\documentclass[reprint, amsmath,amssymb, aps, pra,]{revtex4-2}
\usepackage{graphicx}% Include figure files
\usepackage{physics}
\usepackage{mathtools}
\usepackage{amssymb}
\usepackage{graphicx}
\usepackage{float}
\usepackage[caption=false]{subfig}
\usepackage{svg}
\usepackage{braket}
\usepackage{hyperref}
\usepackage{subfiles}
\usepackage{dcolumn}% Align table columns on decimal point
\usepackage{bm}% bold math
\usepackage{ulem}

\begin{document}
\preprint{APS/123-QED}
\title{Multi-parameter Optimization for Ground-state Cooling of Mechanical Mode using Quantum Dots}

\author{Neelesh Kumar Vij$^{1}$, Meenakshi Khosla$^{2}$, and Shilpi Gupta$^{1,3}$}
\email{corresponding author: ShilpiG@iitk.ac.in}
\affiliation{$^1$Department of Electrical Engineering, Indian Institute of Technology Kanpur, Kanpur-208016, UP, India \\
$^2$Department of Brain and Cognitive Sciences, Massachusetts Institute of Technology, Cambridge, MA-02139, USA \\
$^3$Centre for Lasers and Photonics, Indian Institute of Technology Kanpur, Kanpur-208016, India}
%%\date{\today}

% \preprint{APS/123-QED}
\begin{abstract}
Cooling a mechanical mode to its motional ground state opens up avenues for both scientific and technological advancements in the field of quantum meteorology and information processing. We propose a multi-parameter optimization scheme for ground-state cooling of a mechanical mode using quantum dots. Applying the master equation approach, we formulate the optimization scheme over a broad range of system parameters including detunings, decay rates, pumping rates, and coupling strengths. We implement the optimization scheme on two major types of semiconductor quantum dot systems: colloidal and epitaxial quantum dots. These systems span a broad range of mechanical mode frequencies, coupling rates, and decay rates. Our optimization scheme lowers the steady-state phonon number in all cases by several orders of magnitude. We also calculate the net cooling rate by estimating the phonon decay rate and show that the optimized system parameters also result in efficient cooling. The proposed optimization scheme can be readily extended to other driven systems coupled to a mechanical mode.
\end{abstract}

\maketitle

\section{Introduction}
%Para 1: What is the importance of nanomechanical resonators, why to cool them, and why optimization of cooling process is important
Mechanical resonators have been an essential tool for precision metrology for a long time \cite{Schwab_2005}. With the advancements in nanofabrication techniques, efforts are now oriented towards studying the quantum mechanical aspects of the mechanical resonators \cite{Aspelmeyer_2008, Schwab_2005}. Cooling a mechanical resonator mode to its motional ground state is of particular interest because of wide-ranging applications in the fields of quantum meteorology \cite{Teufel_2009, Krause_2012, Hu_2013, Purdy_2013, Buchmann_2013}, information processing \cite{Wang_2012, Fiore_2011, Schmidt_2012, Stannigel_2012} and testing of quantum-classical boundary \cite{Romero_Isart_2011, Sekatski_2014, Pepper_2012, Blencowe_2013}.

%Para 2-3: What all has been done for cooling oh nanomechanical resonators and what are the challenges?
Coupling of a mechanical resonator mode with a thermal bath leads to heating because of thermalization, while coupling with a dissipative channel leads to cooling. Competing heating and cooling processes decide the extent of cooling of the mechanical resonator mode \cite{Liu_2018}. Therefore, a general approach to cool the resonator mode to its motional ground state is twofold: (a) reduce the heating rate by reducing the coupling of the mechanical resonator mode with the thermal bath \cite{Reinhardt_2017, Norte_2016} and (b) increase the cooling rate by engineering additional dissipative channels with hybrid quantum mechanical systems. Various approaches utilizing solid-state systems such as cooling with color centers \cite{abdi2017spin, kepesidis2013phonon, macquarrie2017cooling, Cortes_2019}, superconducting qubits \cite{PhysRevB.80.144508, PhysRevB.84.094502, PhysRevLett.95.097204}, quantum dots \cite{Zhou_2016, QDCooling2, QDCooling3, Zhu_2012}, and optomechanical systems \cite{PhysRevLett.102.207209, chan2011laser, Liu_2015, Mu_2019, Liu_2017, Guo_2014} have been proposed and studied. To achieve maximum and efficient cooling of the mechanical mode, system parameters that are within experimental control need to be optimized. Many of these parameters depend on the physical realization of the dissipative channel and have been selectively studied in the literature; for example: detuning between the energy levels of the channel and the input laser \cite{kepesidis2013phonon, PhysRevB.80.144508, QDCooling3, Zhou_2016, Mu_2019, Liu_2017, Liu_2015, Guo_2014}, coherent coupling strengths within the channel \cite{abdi2017spin, PhysRevLett.102.207209, chan2011laser, Mu_2019, Zhu_2012, Liu_2017, Liu_2015, Guo_2014}, and the decay rates of the channel \cite{QDCooling3, Zhu_2012, Cortes_2019}. However, simultaneous optimization of multiple system parameters is required to achieve maximum and efficient cooling. 

%Para 4: What is our approach and what does it achieve that has not been done?
Here, we propose a multi-parameter optimization scheme to cool a mechanical mode using a semiconductor quantum dot. Semiconductor quantum dots are one of the most mature solid-state systems \cite{QDreview}. They exhibit tunable optical properties and compatibility with CMOS fabrication technology, allowing for seamless integration into hybrid optomechanical systems at micro- and nano-scales. We formulate an optimization problem over various detunings, decay rates, pumping rates, and coupling strengths of the system using the master equation formalism. We particularly show cooling of a mechanical mode by coupling to two different types of quantum dot systems --- colloidal and epitaxial quantum dots. We show that multi-parameter optimization lowers the steady-state phonon number of the mechanical mode by several orders of magnitude. We also calculate the effective phonon decay rate and show that the optimized system parameters simultaneously result in maximum and efficient cooling.

\section{\label{sec:2} Description of the system}
\subsection{Hamiltonian and master equation}

%Para 6-8: setup of the H, rotated H, master equation
The two widely studied quantum dot systems --- a colloidal quantum dot and an epitaxial quantum dot strongly coupled to an optical cavity --- are routinely modeled as three-level systems \cite{Khosla_2018, Restrepo_2014, UltrafastPolaritons, Biadala}. The three levels in the model for a colloidal quantum dot represent the ground state, the dark exciton, and the bright exciton \cite{Khosla_2018, Biadala}. Similarly, the three levels in the model for an epitaxial quantum dot strongly coupled to an optical cavity represent the ground state, the lower polariton, and the upper polariton \cite{Restrepo_2014, UltrafastPolaritons}. The energy differences between the bright and the dark exciton of a colloidal quantum dot and between the upper polariton and the lower polariton in the strongly coupled epitaxial quantum dot-cavity system are in the regime of mechanical mode energies. Therefore, we model a quantum dot as a three-level system comprising of the ground state \(\ket{0}\), the first excited state \(\ket{1}\), and the second excited state \(\ket{2}\) (Figure \eqref{3ls}). The phonon mode of the mechanical resonator is coupled to the two excited states of the three-level quantum dot system with strength \(g\). We also include a coherent pump of strength \(\Omega\) at frequency \(\omega_p\) between the states \(\ket{0}\) and \(\ket{1}\). The system Hamiltonian can thus be written as (\(\hbar = 1\)):

\begin{figure}
\includegraphics[width = 0.8 \linewidth]{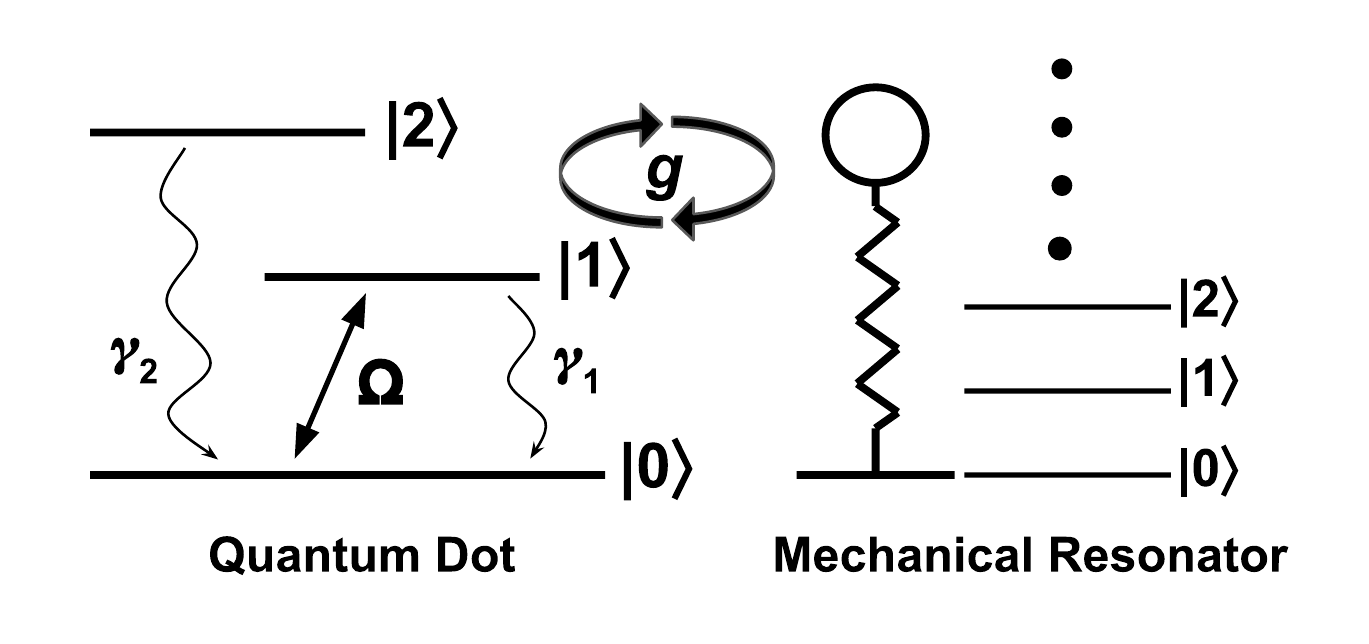}
\caption{A schematic representing a quantum dot coupled to a mode of a mechanical resonator.}
\label{3ls}
\end{figure}

\begin{equation}
\begin{aligned}
\label{non-rotated hamiltonian}
    \mathbf{H_{\text{system}}} &=  \omega_{1}\sigma_{11} + \omega_{2}\sigma_{22} + \omega_{m}b^{\dagger}b + g (\sigma_{12} b^{\dagger} + \sigma_{21}b) \\
    &+ \Omega \big(\sigma_{01}e^{i\omega_{p}t} + \sigma_{10}e^{-i\omega_{p} t} \big)
\end{aligned}
\end{equation}
where \(\omega_1\) and \(\omega_2\) are frequencies of the states \(\ket{1}\) and \(\ket{2}\) respectively, and \(\omega_{m}\) is frequency of the phonon mode. The operator \(\sigma_{ij} = \ket{i}\bra{j}\) represents population operator when \(i = j\) and dipole operator when \(i \neq j\). The annihilation (creation) operator for the phonon mode is \(b\) (\(b^{\dagger}\)). We set the energy of the ground state of the three-level system to zero. To remove the time-dependent terms, we move to a suitable rotated frame and obtain the following rotated Hamiltonian (see derivation in Appendix A):

\begin{equation}
\label{rotated hamiltonian}
\begin{aligned}
    \mathbf{H_{\text{rotated}}} &=  \Delta_1 \sigma_{11} + (\Delta_1 + \Delta_2) \sigma_{22} + \Omega (\sigma_{01} + \sigma_{10}) \\ 
    &+ g (\sigma_{12} b^{\dagger} + \sigma_{21}b)
\end{aligned}
\end{equation}
where \(\Delta_1 = \omega_1 - \omega_p\) and \(\Delta_2 = \omega_2 - \omega_1 - \omega_{m} \). To analyze the complete dynamics of the system, we use the Lindblad master equation for the combined density operator \(\rho\) under the Born-Markov approximation:
 
\begin{equation}
\label{master eq}
    \begin{aligned}
        \dfrac{d \rho}{dt} &= i [\rho, \mathbf{H_{\text{rotated}}}] + \gamma_1 \mathcal{L}[\sigma_{01}] \rho + \gamma_2 \mathcal{L}[\sigma_{02}] \rho \\
        & + \gamma (n_{th} + 1) \mathcal{L}[b] \rho + \gamma n_{th} \mathcal{L}[b^{\dagger}] \rho
    \end{aligned}
\end{equation}
Here, \(\mathcal{L}[\mathcal{O}]\rho = \mathcal{O} \rho \mathcal{O}^{\dagger} - (\mathcal{O}^{\dagger} \mathcal{O}\rho + \rho \mathcal{O}^{\dagger} \mathcal{O})/2\), \(\gamma_1\) and \(\gamma_2\) are the decay rates of \(\ket{1}\) and \(\ket{2}\) respectively, \(\gamma\) is the decay rate of phonon mode, and \(n_{th} = 1/(e^{\omega_{m}/ k_B T} - 1)\) is the phonon number in thermal equilibrium at temperature \(T\) and frequency \(\omega_{m}\). The last two terms in Eq \ref{master eq} account for the coupling of the mechanical mode with the reservoir held at a constant temperature \(T\). 

Our scheme to cool a mechanical mode using a quantum dot can be understood as follows: A coherent pump drives the population between the states \(\ket{0}\) and \(\ket{1}\). The population in the state \(\ket{1}\) transitions to the state \(\ket{2}\) by absorbing a near-resonant phonon from the coupled mechanical mode. The population in the state \(\ket{2}\) then decays to the ground state via the emission of a photon, leading to an overall decrease in the number of phonons in the mechanical mode. We will use Eq \ref{rotated hamiltonian} in the following sub-section for setting up the optimization problem to achieve the minimum phonon number in the mechanical mode in the steady state. 

\subsection{\label{section:optimization_problem} Deriving the optimization problem}

%Para 9: setting up optimization problem for steady state
We write the Heisenberg operator equations for the expectation value of the phonon number operator (\(b^{\dagger}b\)) and the population operator of the second excited state (\(\sigma_{22}\)) using Eq. \ref{rotated hamiltonian}:

\begin{eqnarray}
\begin{aligned}
    \label{Heisenberg}
    \dfrac{d \braket{b^{\dagger}b}}{dt} & = ig \braket{{\sigma_{21} b} - {\sigma_{12} b^{\dagger}}} + \gamma \big( n_{th} - \braket{b^{\dagger}b} \big) \\
    \dfrac{d \braket{\sigma_{22}}}{dt} & = ig \braket{{\sigma_{12} b^{\dagger}} - {\sigma_{21} b}} - \gamma_2 \braket{\sigma_{22}}
\end{aligned}
\end{eqnarray}
Solving the above equations in steady state gives:
\begin{equation}
\label{steadystate_phonon}
    \braket{b^{\dagger}b}_s =  n_{th} - \frac{\gamma_2}{\gamma} \braket{\sigma_{22}}_s
\end{equation}
where the subscript denotes the expectation values calculated in steady state. The above equation brings out the inherent optimization problem present in the combined system. To minimize the steady-state phonon number in the mechanical mode, we need to maximize the term \(\gamma_2 \braket{\sigma_{22}}_s\). However, as we increase \(\gamma_2\), the steady-state population of the state \(\ket{2}\), represented by \(\braket{\sigma_{22}}_s\), decreases and vice-versa. Thus, an optimal value of \(\gamma_2\) exists, which leads to the minimum phonon number. The optimization also depends on the other parameters of the system (decay rates, detunings, coupling strength  and pumping strength) that come into play via the expression of \(\braket{\sigma_{22}}_s\). Next, we analyze the role of each of these parameters in minimizing the steady-state phonon number. 

%Para 10: parameters used
Depending on the platform in which a mechanical mode coupled to a quantum dot is realized, the coupling strengths and the decay rates can range from MHz - GHz \cite{Khosla_2018, Zhou_2016}. For this section, we set \(g = 5\) GHz, \(\omega_2 - \omega_1 = 120.9\) GHz, and an initial temperature of the combined system to \(50\) K, as a generic set of parameters. We also define a figure of merit for cooling of the mechanical mode as \(\mathcal{F} = \braket{b^{\dagger}b}_s/n_{th}\), which needs to be minimized \cite{Cortes_2019}.

\subsubsection{Effect of Detunings}
\begin{figure}
    \includegraphics[width = 0.45\textwidth]{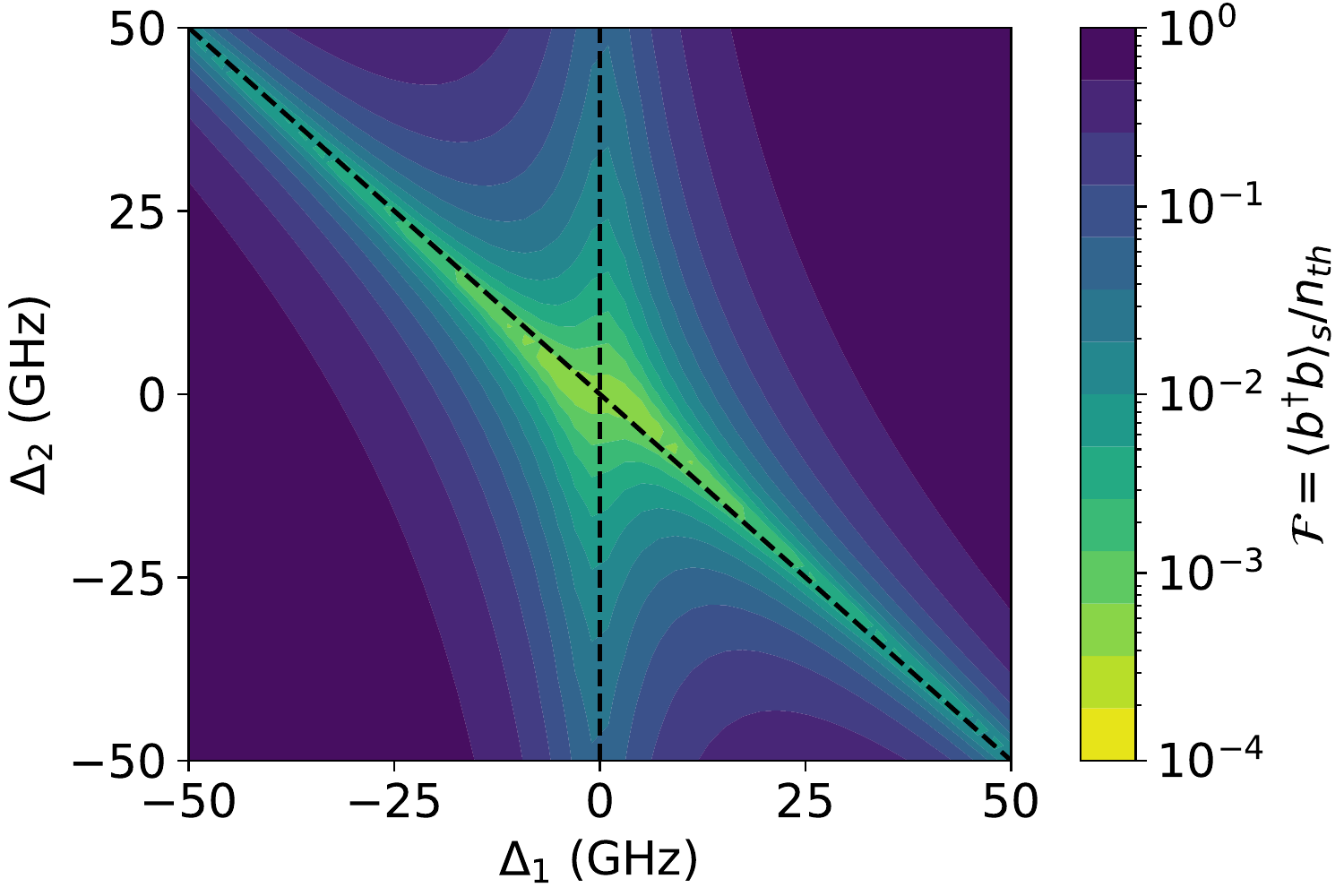}
    \caption{Variation of \(\mathcal{F} = \braket{b^{\dagger}b}_s/n_{th}\) with detunings \(\Delta_1\) and \(\Delta_2\). Here, \(g = 5\) GHz, \(\omega_2 - \omega_1 = 120.9\) GHz, \(\gamma_2 = \Omega = g/2, \gamma_1 = 10^{-1}g, \gamma = 10^{-4}g\), an initial temperature of 50 K, and the cutoff for the Fock state basis \(N = 40\).}
    \label{detuning}
\end{figure}

%Para 11: effect of detuning
First we analyze the effect of the detunings \(\Delta_1\) and \(\Delta_2\) on the figure of merit \(\mathcal{F}\). For this calculation, we fix \(\gamma_2 = \Omega = g/2, \gamma_1 = 10^{-1}g\), and \(\gamma = 10^{-4}g\). Using Eq. \ref{rotated hamiltonian}, \ref{Heisenberg}-\ref{steadystate_phonon} and the quantum toolbox QuTip \cite{JOHANSSON20131234}, we calculate variation of \(\mathcal{F}\) with detunings \(\Delta_1\) and \(\Delta_2\) (Figure \eqref{detuning}). The cutoff for the Fock state basis in the quantum toolbox is set to \(N = 40\). We vary \(\Delta_1\) and \(\Delta_2\) by varying the pump frequency \(\omega_p\) and the phonon mode frequency \(\omega_m\), respectively. As expected, for highly off-resonant interactions, \(\mathcal{F} \approx 1\), implying negligible cooling. Minimization of the figure of merit is achieved along the dashed lines \(\Delta_1 = 0\), which represents resonance between the pump and the \(\ket{0}\) - \(\ket{1}\) transition, and \(\Delta_2 = -\Delta_1\), which represents resonance between the phonon mode and the \(\ket{1}\) - \(\ket{2}\) transition assisted by the pump. Global minimization of \(\mathcal{F}\) happens when both \(\Delta_1 = 0\) and \(\Delta_2 = 0\). However, we note that even when the phonon mode is off-resonant with the \(\ket{1}\) - \(\ket{2}\) transition (i.e. \(\Delta_2 \neq 0)\), a local minimization of \(\mathcal{F}\) can be achieved by tuning the frequency of the coherent pump such that \(\Delta_1 = - \Delta_2\). This offers a convenient way to minimize the phonon number in the case of an off-resonant phonon mode.

%Para 12: effect of detuning: appendix results
The optimal values of the detunings \(\Delta_1\) and \(\Delta_2\) are also mandated by the relative strengths of \(g\) and \(\Omega\). When \(g\) and \(\Omega\) are not of the same order, the minimization of \(\mathcal{F}\) happens for non-zero detuning values (see derivation in Appendix B1). Since the pumping strength \(\Omega\) is a control parameter in experiments, we choose it to be of the same order as \(g\): \(\Omega \approx g\) and set \(\Delta_1 = \Delta_2 = 0\) in further calculations. We note that the optimal values of the detunings are governed by only the system Hamiltonian (Eq \ref{rotated hamiltonian}), and therefore, they are invariant to the changes in the decay rates.

\subsubsection{Effect of Decay Rates}
\begin{figure}
    \centering
    \includegraphics[width = 0.5\textwidth]{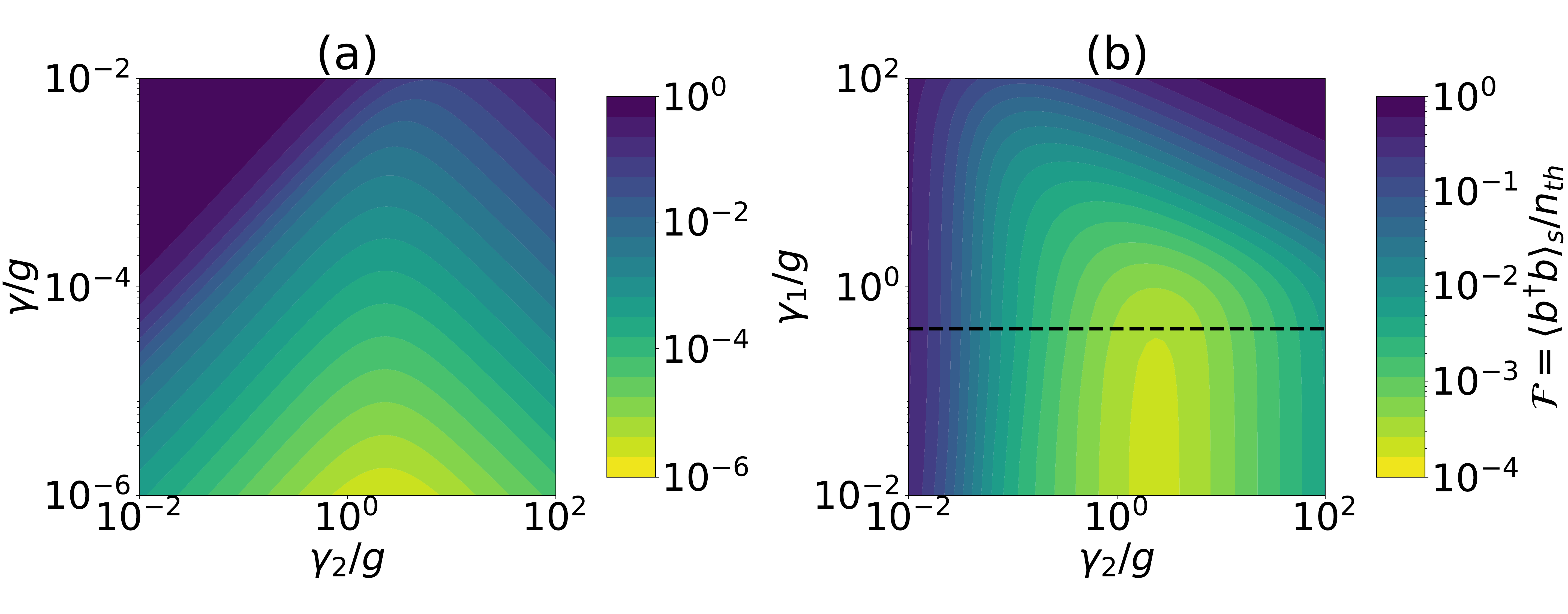}
    \caption{Variation of \(\mathcal{F}\) with (a) phonon decay rate \(\gamma\) and decay rate of the second excited state \(\gamma_2\) for \(\gamma_1 = 10^{-1}g\) (b) decay rate of the first excited state \(\gamma_1\) and the second excited state \(\gamma_2\) for \(\gamma = 10^{-4}g\). The black dashed horizontal line denotes \(\gamma_1 = \Omega\). Here, \(g = 5\) GHz, \(\Delta_1 = \Delta_2 = 0\), \(\omega_2 - \omega_1 = 120.9\) GHz, \(\Omega = g/2\), an initial temperature of 50 K, and the cutoff for the Fock state basis \(N = 40\).}
    \label{decay_rates_plot}
\end{figure}

%Para 13: effect of decay rates: gamma and gamma2
Next, we analyze the effect of the decay rates on the figure of merit \(\mathcal{F}\). Figure \ref{decay_rates_plot}a plots the variation of \(\mathcal{F}\) as a function of the phonon decay rate \(\gamma\) and the decay rate of the second excited state \(\gamma_2\). We fix \(\Omega = g/2\) and  \(\gamma_1 = 10^{-1}g\), and vary \(\gamma\) and \(\gamma_2\) over a large range of values spanning across four orders of magnitude. We make the following observations:
\begin{itemize}
    \item[--] \(\mathcal{F}\) decreases as the phonon decay rate \(\gamma\) decreases. A lower \(\gamma\) reduces the interaction of the mechanical mode with the phonon reservoir, reducing the heating rate of the mechanical mode. For further calculations, we set \(\gamma \ll g\).
    \item[--] There exists an optimal decay rate of the second excited state \(\gamma_2\) that minimizes \(\mathcal{F}\). This observation is consistent with our explanation presented in the beginning of this section using Eq. \ref{steadystate_phonon}.
\end{itemize}

%Para 14: effect of decay rates: gamma1 
Figure \ref{decay_rates_plot}b shows the variation of \(\mathcal{F}\) with \(\gamma_1\) and \(\gamma_2\) for \(\Omega = g/2\) and \(\gamma = 10^{-4}g\). The black dashed line represents the condition \(\gamma_1 = \Omega\). Consistent with Figure \ref{decay_rates_plot}a, Figure \ref{decay_rates_plot}b exhibits the same optimal \(\gamma_2\) when \(\gamma_1 \lesssim \Omega\), which leads to a globally minimized \(\mathcal{F}\). When \(\gamma_1 \gtrsim \Omega\), decoherence sets in between the states \(\ket{0}\) and \(\ket{1}\) requiring a lower value for optimal \(\gamma_2\) and increasing the minimum achievable value of \(\mathcal{F}\). When \(\gamma_1 > \Omega\) and \(\gamma_2\) is greater than its optimal value, decoherence dominates and leads to \(\mathcal{F} \approx 1\) as depicted in the upper right region of the plot. We note that when \(\Omega \not\approx g\), \(\mathcal{F}\) can be minimized either by introducing non-zero detunings (see Appendix B1) or by varying the decay rates (see Appendix B2), both of which lead to an increase in the minimum achievable \(\mathcal{F}\).

%Para 15: Overall conclusion of the above qutip exercise
The above discussion gives us a broad set of conditions for the system parameters which leads to minimization of the figure of merit \(\mathcal{F}\): (i) \(\Omega \approx g\) with \(\Delta_1 = \Delta_2 = 0\), (ii) \(\gamma \ll g\), (iii) \(\gamma_1 \lesssim \Omega\), and (iv) an optimal value for \(\gamma_2\). To perform further optimization, we identify the following two regimes of decay rates exemplified by quantum dot systems: \(\gamma_1 \ll \gamma_2\) and \(\gamma_1 = \gamma_2\). In the next section, we analytically formulate the cooling optimization problem for these two regimes and discuss the outcomes for the relevant quantum dot systems.

\section{\label{sec3}cooling optimization for different regimes}
%Para 16: Specifying notation for states of the combined system
In this section, we derive steady-state analytical models for the two regimes of the decay rates by restricting the excitations of the mechanical resonator phonon mode to one. The approximation is valid because we characterize the mechanical resonator mode in steady state, which is achieved when the phonon mode is cooled close to the ground state, and thus the probability of higher excitations is negligible. We compare the results of our approximated analytical model with numerical simulations performed for a sufficiently large basis of phonon mode to ensure numerical stability. Through the comparison, we show that the approximation is indeed valid and later discuss its limitations. Furthermore, we calculate the effective phonon decay rate and show that the parameters that minimize the figure of merit also result in a large cooling rate.

\subsection{Regime: \(\gamma_1 \ll \gamma_2\) \label{section g1 << g2}}
%Para 17: gamma1 << gamma2 regime: setting parameters of colloidal QD

We derive the equations of motion for the density matrix elements of the combined system (see Appendix C) and solve them under the approximation \(\gamma_1 \ll \gamma_2\) to obtain the expression for the steady state phonon number (see Appendix C1). To set values of the parameters for this regime, we consider a system of a colloidal quantum dot, specifically a cadmium selenide quantum dot coupled to its confined phonon mode via deformation potential. For temperatures \(< 20\) K, a colloidal quantum dot can be approximated as a three-level system with the first and the second excited states \(\ket{1}\) and \(\ket{2}\) being the dark and the bright states, respectively \cite{Khosla_2018, Biadala}. The dark state of the colloidal quantum dot has a lifetime of \(\sim\) 1 ms and the bright state has a lifetime of \(\sim\) 10 ns. Therefore, this colloidal quantum dot - phonon system exemplifies the regime of \(\gamma_1 \ll \gamma_2\). To facilitate coherent pumping to the dark state, a two-photon absorption technique can be used \cite{two-photon}. Consistent with the literature, we set the following values for the system parameters: \(\omega_m = 241.8\) GHz, \(g = 20\) GHz, \(\gamma_1 = 10^{-6}\) GHz, \(\gamma = 10^{-3}\) GHz, an initial temperature of \(17\) K, and the cutoff for the Fock state to be \(N = 40\) for further calculations in this subsection \cite{Khosla_2018}.

\begin{figure}[h]
    \centering
    \includegraphics[width = 0.49\textwidth]{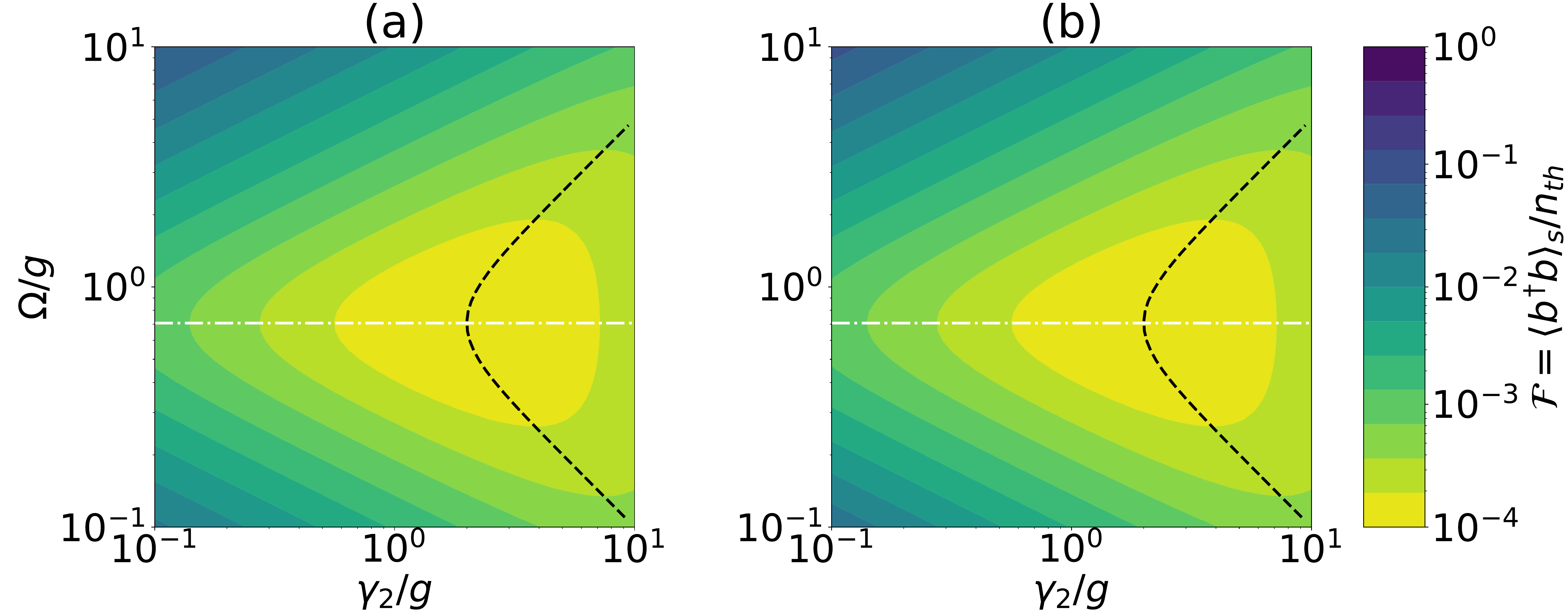}
    \caption{\textbf{Regime \(\gamma_1 \ll \gamma_2\)}: Variation of \(\mathcal{F}\) with pumping strength \(\Omega\) and decay rate of second excited state \(\gamma_2\) using (a) approximated analytical model (Eq C.2) and (b) exact simulation model. The white dot-dashed line and the black dashed curve represent the optimal parameters \(\Omega_o\) and \(\gamma_{2_o}\), respectively, as derived in Eq \eqref{optimal g1 << g2}. Here, \(\omega_m = 241.8\) GHz, \(g = 20\) GHz, \(\gamma_1 = 10^{-6}\) GHz, \(\gamma = 10^{-3}\) GHz, an initial temperature of \(17\) K, and the cutoff for the Fock state basis \(N = 40\).}
    \label{optimal plot g1 << g2}
\end{figure} 

%Para 18: gamma1 << gamma2 regime: discuss fig 4a and 4b
We treat the strength of the coherent pumping \(\Omega\) and the decay rate of the bright state \(\gamma_2\) as the control parameters in this system. The decay rate of the bright state can be altered via Purcell enhancement by coupling to an optical cavity. Purcell factors upto \(10^{4}\) have been achieved using appropriate cavity geometry \cite{Lu_2017, Lee_2015}. Therefore, we plot the figure of merit \(\mathcal{F}\) as a function of \(\Omega\) and \(\gamma_2\) using the expression of \(\braket{b^{\dagger}b}_s\) (Appendix C1) in Figure \ref{optimal plot g1 << g2}a. We also exactly calculate \(\mathcal{F}\) by numerically solving the master equation (Eq \eqref{master eq}) using the quantum toolbox without restricting the mechanical mode excitations (Figure \ref{optimal plot g1 << g2}b). Both Figures \ref{optimal plot g1 << g2}a and \ref{optimal plot g1 << g2}b suggest that \(\mathcal{F}\) minimizes for certain values of \(\Omega\) and \(\gamma_2\), which we next calculate analytically. Since \(\gamma_2 \gg \gamma\), we neglect the higher order terms of \(\gamma\) in the expression of \(\braket{b^{\dagger}b}_s\) (Appendix C1) and arrive at the expressions for optimal pumping strength \(\Omega_o\) and optimal decay rate of the second excited state \(\gamma_{2o}\):

\begin{equation}
\label{optimal g1 << g2}
\begin{aligned}
    \Omega_o &= \dfrac{g}{\sqrt{2}} \\
    \gamma_{2_o} &= \sqrt{4\Omega^2 + \dfrac{g^4}{\Omega^2}}
\end{aligned}
\end{equation}
%Para 19: gamma1 << gamma2 regime: plotting and discussion of optimal expressions
We plot \(\Omega_o/g\) as a white dot-dashed line and \(\gamma_{2_o}/g\) as a black dashed curve in Figure \ref{optimal plot g1 << g2}. We observe that the optimal pumping strength is a function of only the coupling strength while the optimal decay rate of the second excited state is a function of both the pumping strength and the coupling strength. \(\Omega_o\) depends on only the coupling strength because the decay rate of the first excited state \(\gamma_1\) is negligible compared to the other rates in this system. On the other hand, \(\gamma_{2_o}\) depends on \(\Omega\) and \(g\) because the two coherent processes compete with the incoherent process for population transfer to and from the second excited state. We further include the effect of pure dephasing in our model using the quantum toolbox (see Appendix D) and observe that \(\mathcal{F}\) and the optimal values of the pumping strength and the decay rate of the second excited state (Eq \ref{optimal g1 << g2}) show negligible variation with the inclusion of pure dephasing rates observed in the literature \cite{Khosla_2018}.

\begin{figure*}[htb!]
    \centering
    \includegraphics[width = \textwidth]{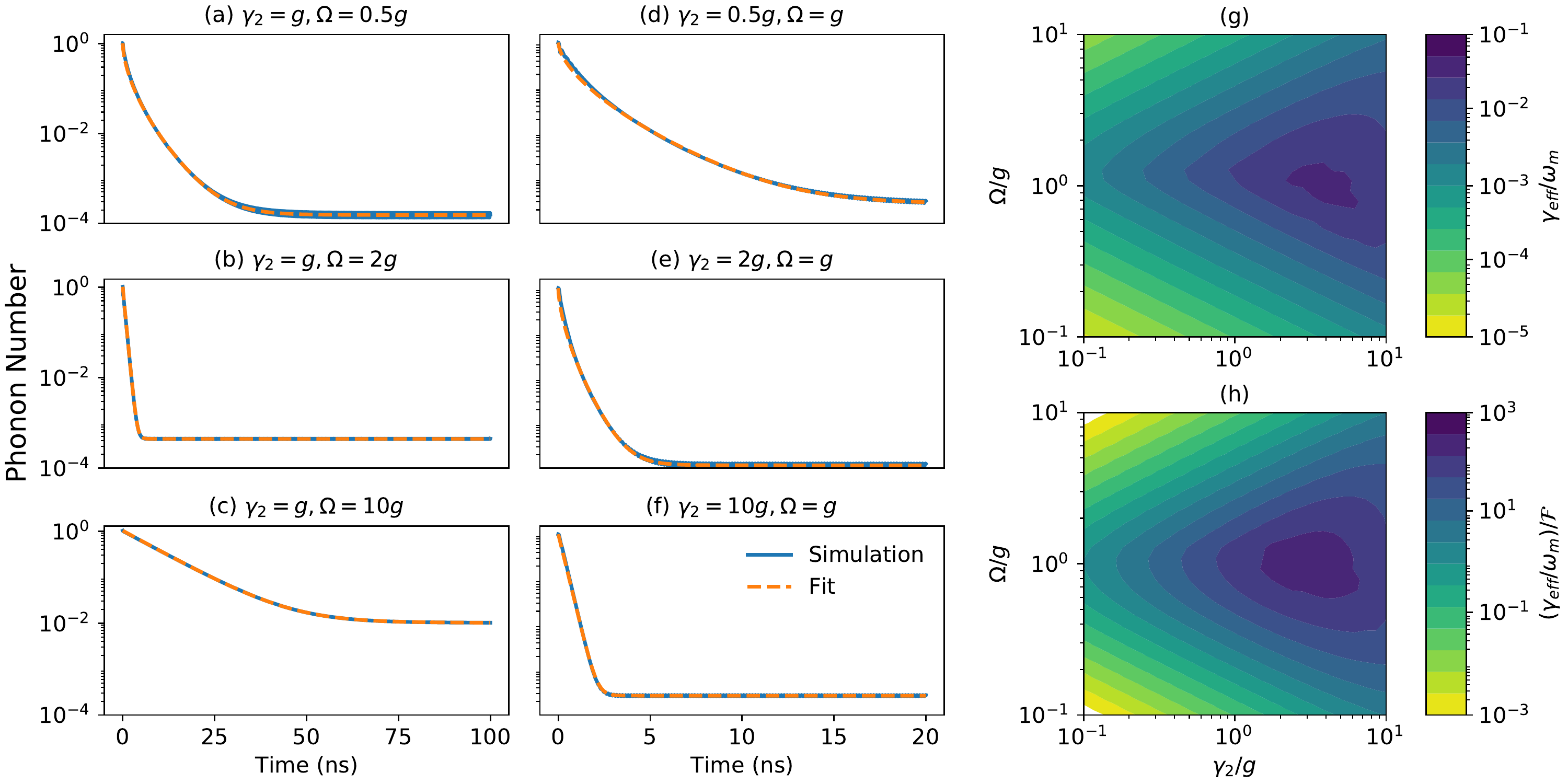}
    \caption{\textbf{Regime \(\gamma_1 \ll \gamma_2\)}: (a)-(f) Time-dependent decay of average phonon number of the mechanical mode and the corresponding stretched-exponential fit, (g) \(\gamma_{\text{eff}}/\omega_m\), (h) \((\gamma_{\text{eff}}/\omega_m)/\mathcal{F}\), over a range of values of \(\Omega\) and \(\gamma_2\). Here, \(\omega_m = 241.8\) GHz, \(g = 20\) GHz, \(\gamma_1 = 10^{-6}\) GHz, \(\gamma = 10^{-3}\) GHz, an initial temperature of \(17\) K, and the cutoff for the Fock state basis \(N = 40\).}
    \label{decay rate g1<<g2}
\end{figure*}

% Para 20: When the theoretical model breaks down
The theoretical and simulation results are in good agreement, justifying that the higher order phonon excitations are sparsely populated and thus can be neglected. The approximation however breaks down if the rate of bulk phonon decaying into the system \(( = \gamma n_{th})\) is comparable to the coupling strength \(g\). This can happen when either \(n_{th}\), governed by the initial temperature, or \(\gamma\), governed by the quality factor of the mechanical mode, is large. We note that the choice of physical system (quantum dot here) dictates the initial value of \(n_{th}\) and the frequency of the mechanical mode. Our model can be readily extended to any other physical system obeying the approximations laid out until now. 

%Para 20: gamma1 << gamma2 regime: discuss fig 5
\textit{Cooling rate}: With the optimization in place, we now estimate the rate of the cooling process by calculating the effective phonon decay rate (\(\gamma_{\text{eff}}\)). As the decay of a general mechanical state may involve simultaneous decay of multiple phonon Fock states, it will be a multi-exponential decay process. In the literature, the two prominent models used for multi-exponential decay processes are the stretched exponential fit \cite{StretchedExponentialFit} and the log-normal fit \cite{LogNormalFit}. Here, we employ the stretched exponential fit. We plot the variation of phonon number with time (Figure \ref{decay rate g1<<g2}a-f), using the quantum toolbox, and fit it to a stretched exponential of the form:

\begin{equation}
    \braket{b^{\dagger}b}(t) = (n_{th} - \braket{b^{\dagger}b}_s) e^{-(\gamma_{\text{eff}}t)^{\beta}} + \braket{b^{\dagger}b}_s
    \label{eq:extended_exp}
\end{equation}
where \(\braket{b^{\dagger}b}_s\) is the steady state phonon number and \(\beta \in (0,1]\) is the stretch parameter with \(\beta = 1\) representing a single exponential decay function. The effective phonon decay rate depends on the initial condition of the mechanical mode. Here, we assume that initially the quantum dot is in the ground state and the mechanical mode is in the thermal state. Repeating the procedure of fitting and extracting the effective phonon decay rate, \(\gamma_{\text{eff}}\), over a range of values of \(\Omega\) and \(\gamma_2\), we plot \(\gamma_{\text{eff}}/\omega_m\) in Figure \ref{decay rate g1<<g2}g. From our simulations, we observe that the phonon-decay process is indeed muli-exponential because the extracted values of \(\beta\) for the fits \(\in [0.6, 1)\). The mean absolute percentage error for all the fits is \(< 5 \%\). We observe that the same range of \(\Omega\) and \(\gamma_2\) maximize \(\gamma_{\text{eff}}\) (Figure \ref{decay rate g1<<g2}g) and minimize \(\mathcal{F}\) (Figure \ref{optimal plot g1 << g2}). To demonstrate this point clearly, we plot the ratio of \(\gamma_{\text{eff}}/\omega_m\) to  \(\mathcal{F}\) in Figure \ref{decay rate g1<<g2}h. Large value of this ratio within the optimized ranges of \(\Omega\) and \(\gamma_{2}\) implies that our optimization process simultaneously ensures maximum and efficient cooling of the mechanical mode.

\subsection{Regime: \(\gamma_1 = \gamma_2\)}
%Para 21: gamma1 = gamma2 regime: discuss the polariton system
For \(\gamma_1 = \gamma_2\) regime, we first consider an optical cavity mode strongly coupled to an epitaxial quantum dot, which is modeled as a two-level system (TLS), forming polaritons that in turn couple to a mode of a mechanical resonator. The Hamiltonian for the system takes the following form (details in Appendix E) \cite{Restrepo_2014}:

\begin{equation}
    \begin{aligned}
        \mathbf{H_{\text{system}}} &= \omega_{1}\sigma_{11} + \omega_{2} \sigma_{22} + \omega_m b^{\dagger} b - \Omega \sin{\theta} \big( \sigma_{01}e^{i\omega_pt} \\
        &+ \sigma_{10}e^{-i\omega_pt} \big) + g \sin \theta \cos \theta (\sigma_{12} b + \sigma_{21} b^{\dagger})
    \end{aligned}
    \label{hamiltonian g1=g2}
\end{equation}
where \(\ket{1}\) and \(\ket{2}\) are the two polariton states of the cavity-TLS system with frequencies \(\omega_1\) and \(\omega_2\) respectively, \(\ket{0}\) is the ground state, \(g\) is the coupling strength between the polariton states and the mechanical resonator mode, and \(\tan(2\theta) = 2G/\Delta\) with \(G\) and \(\Delta\) as the coupling strength and detuning, respectively, between the TLS and the cavity mode. We set the detuning between the TLS and cavity mode to zero (\(\theta = \pi/4\)), which results in the decay rates, \(\gamma_1\) and \(\gamma_2\), of the polariton states to be equal \cite{Neuman:18}. The decay rates \(\gamma_1\) and \(\gamma_2\) are functions of both the quantum dot and the cavity decay rates. We use \(\gamma_2\) to denote both the decay rates of this system in further discussions. With this simplification, the Hamiltonian in Eq \ref{hamiltonian g1=g2} is equivalent to the Hamiltonian in Eq \ref{non-rotated hamiltonian}. Therefore the system follows the same optimization rules on pumping strength, detunings, and phonon decay rate as laid out in section \ref{section:optimization_problem}. For further optimization, we solve the equations of motion for the density matrix elements of the combined system (Appendix C) to obtain the expression for the steady state phonon number under the condition \(\gamma_1 = \gamma_2\) (Appendix C2).

\begin{figure}[h]
    \centering
    \includegraphics[width = 0.5\textwidth]{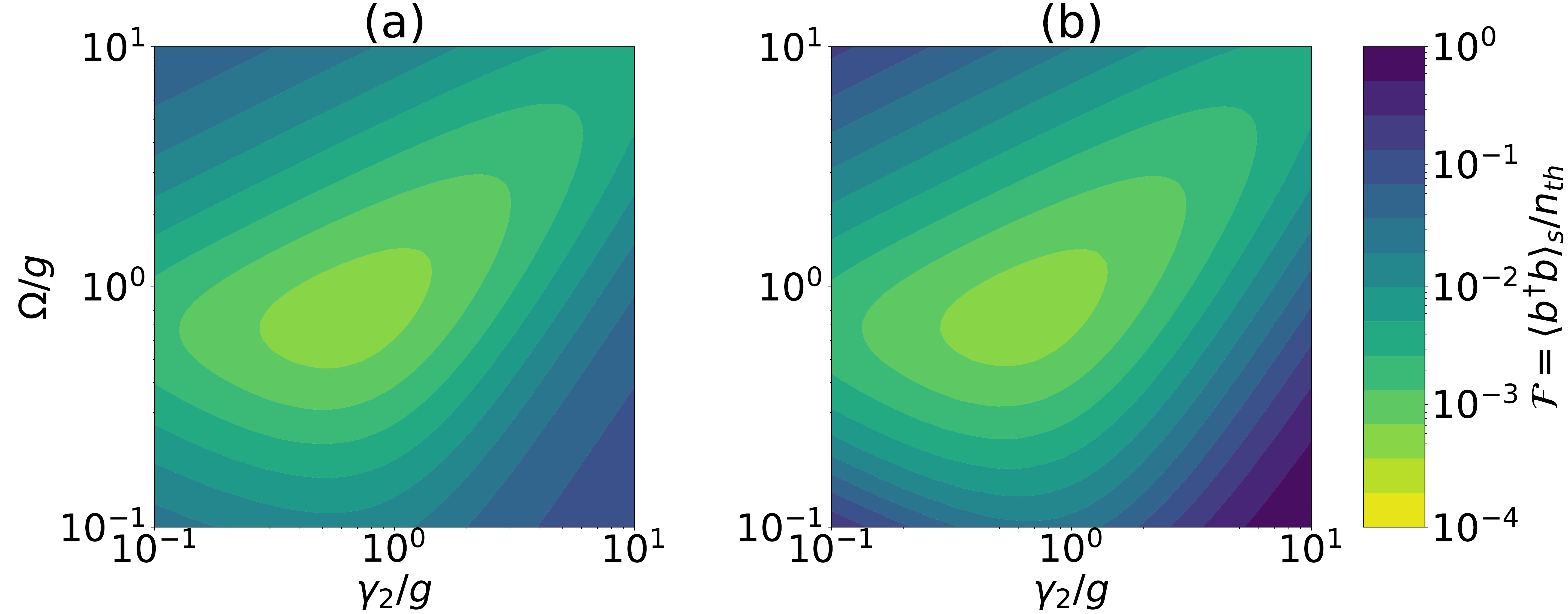}
    \caption{\textbf{Regime \(\gamma_1 = \gamma_2\)}: Variation of \(\mathcal{F}\) with \(\Omega\) and \(\gamma_2\) using (a) approximated analytical model (Eq C.3) and (b) exact simulation model. Here, \(G = 5\) GHz, \(g = 0.002G = 10\) MHz, \(\omega_m = 2G, \gamma =  \omega_m/Q_m = 10^{-7}\) GHz, an initial temperature of \(2.63\) K corresponding to \(n_{th} = 5\), and the cutoff for the phonon Fock state \(N = 40\).}
    \label{optimal plot g1=g2}
\end{figure}

\begin{figure*}[htb!]
    \centering
    \includegraphics[width = \textwidth]{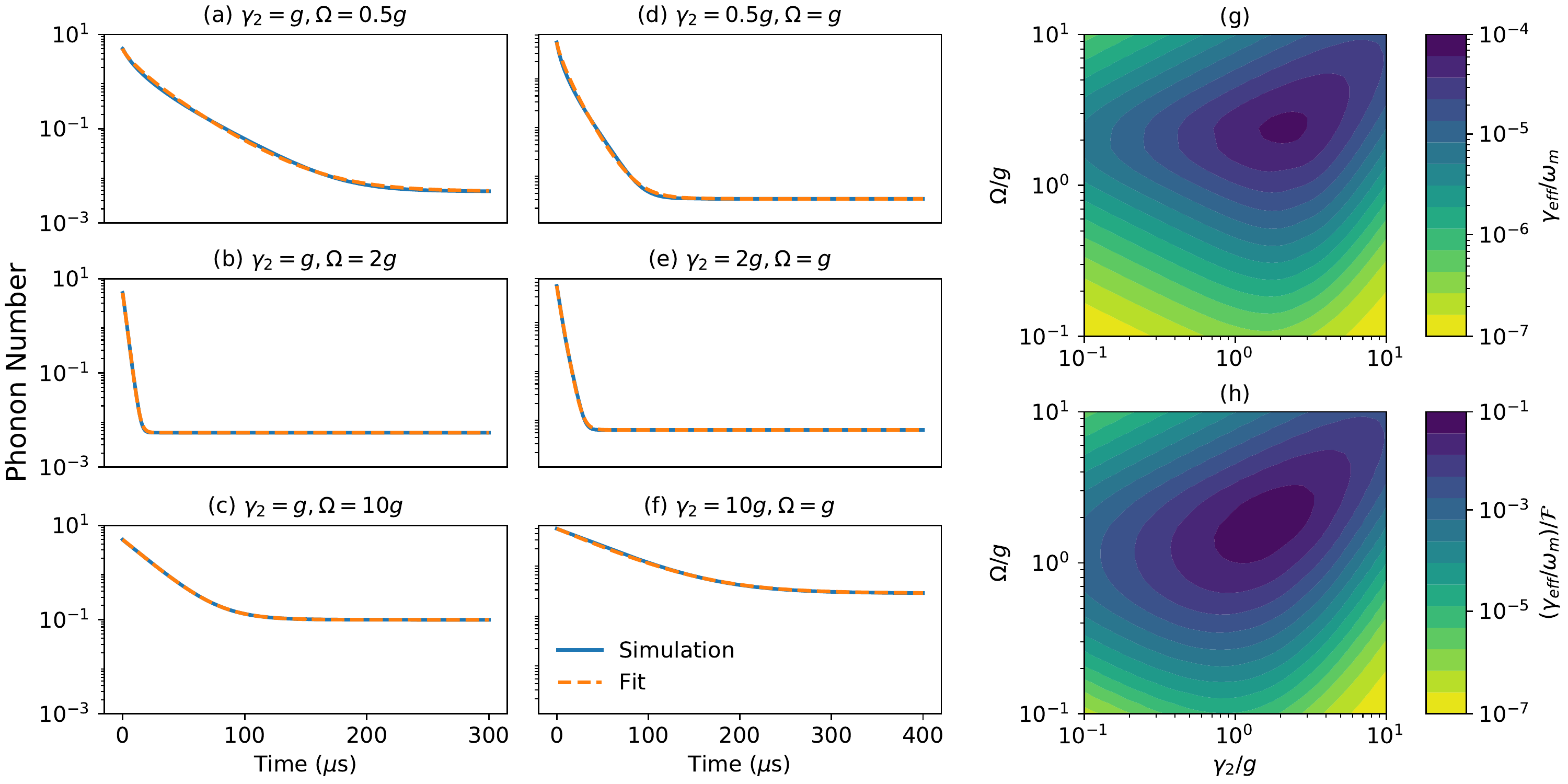}
    \caption{\textbf{Regime \(\gamma_1 = \gamma_2\)}: (a)-(f) Time-dependent decay of average phonon number of the mechanical mode and the corresponding stretched-exponential fit, (g) \(\gamma_{\text{eff}}/\omega_m\), (h) \((\gamma_{\text{eff}}/\omega_m)/\mathcal{F}\), over a range of values of \(\Omega\) and \(\gamma_2\). Here, \(G = 5\) GHz, \(g = 0.002G = 10\) MHz, \(\omega_m = 2G, \gamma =  \omega_m/Q_m = 10^{-7}\) GHz, an initial temperature of \(2.63\) K corresponding to \(n_{th} = 5\), and the cutoff for the phonon Fock state \(N = 40\).}
    \label{phonon_decay_g1=g2}
\end{figure*}

%Para 22: gamma1 = gamma2 regime: parameter values
To set the values of the parameters, we consider a InGaAs quantum dot coupled to a GaAs microdisk cavity. It has been demonstrated that microdisk cavities can support GHz-frequency mechanical modes with mechanical quality factors upto \(10^9\) \cite{high-Q-microdisk}, optical quality factors upto \(6 \times 10^6\)  \cite{microdisc_Q_factor}, and optomechanical coupling strengths in the kHz - MHz range \cite{optomechanical_coupling_number}. The coupling strength between the quantum dot and cavity ranges from a few GHz to \(50\) GHz at low temperatures \cite{microdisk-QD-review-article, microdisk-QD_coupling}. Here, we set \(G = 5\) GHz, \(g = 0.002 G = 10\) MHz, \(\omega_m = 2 G, \gamma =  \omega_m/Q_m = 10^{-7}\) GHz, and an initial temperature of \(2.63\) K corresponding to \(n_{th} = 5\). We choose a small initial temperature to keep computational complexity, associated with a large \(n_{th}\) in quantum toolbox simulations, manageable.

We treat the strength of coherent pumping \(\Omega\) and the decay rate of the polariton states \(\gamma_2\) \((=\gamma_1)\) as the control parameters. The decay rate \(\gamma_2\) can be enhanced or suppressed by engineering the local density of optical states \cite{Lu_2017, Lee_2015, lodahl2004controlling, UltrafastPolaritons}. Therefore, we plot the figure of merit \(\mathcal{F}\) as a function of \(\Omega\) and \(\gamma_2\) in Figure \ref{optimal plot g1=g2}a using the analytically derived expression (Eq. G.1) and in Figure \ref{optimal plot g1=g2}b by numerically solving the master equation (Eq. \ref{master eq} and \ref{hamiltonian g1=g2}) using quantum toolbox without the approximations made for the analytical calculations. We observe that the optimal value of \(\Omega\) (\(\gamma_2\)) is a function of the optomechanical coupling strength and \(\gamma_2\) (\(\Omega\)). Furthermore, the optimal values of \(\Omega\) and \(\gamma_2\) are of the order of the optomechanical coupling strength \(g\), as also observed in the colloidal quantum dot case in Section \ref{section g1 << g2}, and we again attribute this nature to the competing dynamics of the three rates for population transfer to and from the upper polariton state. Obtaining analytical expressions for optimal \(\Omega\) and \(\gamma_2\) is difficult because of the sheer complexity of the expression of the steady state phonon number (Eq. C.3). However, they can be obtained by fitting appropriate functions to the calculated data.

In the current state-of-the-art microdisk cavities, the decay rate of the cavity mode is of the order of 100 MHz \cite{microdisc_Q_factor} and the decay rate of the quantum dot is of the order of MHz \cite{HighCooperativity}. Therefore, the decay rate of the polariton is dominated by the decay rate of the cavity. For the polariton coupled mechanical resonator mode case, our optimization scheme requires suppressing the decay rate \(\gamma_2\) because the optomechanical coupling strength is in the MHz regime. Since the minimum achievable \(\gamma_2\) is currently limited by the cavity decay rate, further improvements in cavity design and fabrication can enable higher optical quality factors, and thereby, the optimal value of \(\gamma_2\).

%Para 23: gamma1 = gamma2 regime: effect of pure dephasing and incoherent pumping
We also include the effect of pure dephasing observed in this quantum dot system \cite{HighCooperativity} and notice that the additional decoherence reduces the net achievable cooling and shifts the optimized parameter range (see Appendix F1). As opposed to CdSe quantum dots (\(\gamma_1 \ll \gamma_2\) regime), the dephasing rate for InGaAs quantum dots (\(\gamma_1 = \gamma_2\) regime) reported in the literature is larger than the coupling strength \(g\). Therefore, the optimized parameter range shifts in order to account for the additional decoherence. We further include a small above-band incoherent pump to the polaritonic states, consistent with experiments probing polaritons \cite{Polariton_incoherent_pumping_experiment}, and observe similar changes (see Appendix F2).

%Para 24: gamma1 = gamma2 regime: discuss fig 7 
\textit{Cooling rate}: With the optimization in place, we now proceed to calculate the effective decay rate for the mechanical resonator mode. We choose the same fit function as in Eq \ref{eq:extended_exp} and a similar initial conditions: ground state for the polariton system and thermal state for the mechanical resonator mode. Figures \ref{phonon_decay_g1=g2}a-f show variation of phonon number with time for different values of \(\Omega\) and \(\gamma_2\), and the associated fits. The extracted value of \(\beta\) for the fits \(\in [0.7, 1)\), thus revealing the presence of a multi-exponential decay. The mean absolute percentage error \(< 5\%\) for all the fits. Figures \ref{phonon_decay_g1=g2}g shows dependence of \(\gamma_{\text{eff}}/\omega_m\) over a range of values of \(\Omega\) and \(\gamma_2\). We also plot the ratio of \(\gamma_{\text{eff}}/\omega_m\) to \(\mathcal{F}\) in Figure \ref{phonon_decay_g1=g2}h. Similar to the \(\gamma_1 \ll \gamma_2\) regime, a large value of the ratio simultaneously leads to maximum and efficient cooling of the mechanical mode.

% Para 25: Mn doped QD system
Another system that satisfies the condition \(\gamma_1 = \gamma_2\) is a InAs/GaAs quantum dot doped with a single Manganese atom. At low temperatures \((\sim 4\) K), the excitations of the higher energy states can be neglected and the system is approximated as a four-level system, comprising of two ground states and two excited states, in the absence of magnetic field \cite{OlivierMnQD, MnDopedQD}. In line with the cooling procedure described in Section II, we assume that the two excited states couple to a mechanical resonator mode and the first excited state is pumped coherently from the lowest energy ground state. This leaves the higher energy ground state decoupled from the cooling process. We show that the four-level system can be approximated as a three-level system for permitted value of system parameters, and therefore, follows the optimization method outlined in this subsection and cools the coupled mechanical mode (Appendix G).

\section{\label{conc} Conclusion and Discussion}
In conclusion, we proposed cooling of a mechanical resonator mode using quantum dots. We formulated an optimization problem, using the master equation approach, over a broad range of system parameters including detunings, decay rates, coupling strengths, and pumping rates. Through particular examples of two quantum dot systems --- colloidal and epitaxial quantum dots --- with mechanical mode frequencies ranging from MHz to GHz, we showed that ground-state cooling of the mechanical mode is achieved by optimizing the system parameters. We also calculated the cooling rate by estimating the rate of phonon decay and showed that the optimized system parameters simultaneously result in both maximum and efficient cooling. We note that the cooling is robust to small variations in the optimized system parameters because of delocalized maxima observed in Figures \ref{decay rate g1<<g2}h and \ref{phonon_decay_g1=g2}h.

While some previous works have looked at selective optimization of system parameters, our model provides optimization over a broad range of multiple parameters. For example, our model is a generalization to the model presented in \cite{Cortes_2019} where the three-level system was effectively reduced to a two-level system on application of a strong incoherent pump to the first excited state. Replacing the coherent pump with an incoherent pump, our model reproduces the result of \cite{Cortes_2019} for the case of a single atom, giving the optimal decay rate \(\gamma_{2_o} = 2g\) (Appendix H). Accounting for experimental limitations, our work provides a generalized framework for optimizing ground-state cooling of a mechanical resonator mode using a quantum dot. 

\section*{Acknowledgments}
We acknowledge funding support from Indo-French Centre for the Promotion of Advanced Research – CEFIPRA (Project No. 64T3-2). We thank Harshawardhan Wanare, Olivier Krebs, Saikat Ghosh, and Shruti Shukla for insightful discussions.

\appendix
\renewcommand{\theequation}{\thesection.\arabic{equation}}
\renewcommand{\thesubsection}{\thesection \arabic{subsection}}

\section{\label{Appendix:Rot_H} Conversion of Hamiltonian to a rotated frame}
Using a unitary transformation, defined by \( \displaystyle \mathbf{U} = e^{-i \mathbf{H_{o}} t/ \hbar}\), we transform our system Hamiltonian \(\mathbf{H_{\text{system}}}\) to \(\mathbf{H_{\text{rotated}}} = \mathbf{U^{\dagger}} \mathbf{H_{\text{system}}} \mathbf{U} - \mathbf{H_{o}}\) into a new rotating frame of reference \cite{scully}. We choose \(\mathbf{H_{o}} = \alpha \sigma_{11} + \beta \sigma_{22} + \zeta b^{\dagger}b\) and determine \(\alpha = \omega_p, \beta = \omega_p + \omega_m ,\) and \(\zeta = \omega_m\) such that the rotated Hamiltonian takes the following time-independent form (Eq. \ref{rotated hamiltonian}): 
  
\begin{equation}
\label{eq:Appendix_rotatedH}
    \begin{aligned}
        \mathbf{H_{\text{rotated}}} &=  \Delta_1 \sigma_{11} +  (\Delta_1 + \Delta_2) \sigma_{22}  + \Omega (\sigma_{01} + \sigma_{10})\\
        & +  g (\sigma_{12} b^{\dagger} + \sigma_{21}b)
    \end{aligned}
\end{equation}
where \(\Delta_1 = \omega_1 - \omega_p\) and \(\Delta_2 = \omega_2 - \omega_1 - \omega_{m} \). \\

\section{\label{Appendix:detuning_opt} Variation of \(\mathcal{F}\) in the regime \(\Omega \not\approx g\)}
Here, we discuss the variation of \(\mathcal{F}\) with detunings and decay rates for two other cases: \(\Omega \gg g\) and \(g \gg \Omega\).

\subsection{Detunings}
In the regime \(\Omega \gg g\) or \(g \gg \Omega\), we neglect the non-dominant coherent process in the rotated Hamiltonian and subsequently calculate the eigenenergies and the modified detunings. To obtain the optimal detuning conditions, these modified detunings should be set to zero for resonant interaction.

When \(\Omega \gg g\), the eigenenergies of the rotated Hamiltonian, neglecting the Jaynes-Cummings phonon coupling term are: \(\omega_{\pm} = \Delta_1/2 \pm \sqrt{\Delta_1^2/4 + \Omega^2}, \omega_2 = \Delta_1 + \Delta_2\). The second excited state \(\ket{2}\) now couples to both the dressed states. The modified detunings are \(\Delta_{\pm} = \omega_2 - \omega_{\pm} = \Delta_2 + \Delta_1/2 \mp \sqrt{\Delta_1^2/4 + \Omega^2}\). Setting them to zero gives the optimal detuning condition:

\begin{equation}
    \Delta_2 = -\Delta_1/2 \pm \sqrt{\Delta_1^2/4 + \Omega^2}
    \label{optimal_detuning_pump>>g}
\end{equation}

For the case \(g \gg \Omega\), neglecting the pumping term and calculating the eigenenergies gives: \(\omega_o = 0, \omega_{\pm} = \Delta_1 + \Delta_2/2 \pm \sqrt{\Delta_2^2/4 + g^2}\). The modified detunings \(\Delta_{\pm} = \omega_{\pm} - \omega_o = \Delta_1 + \Delta_2/2 \pm \sqrt{\Delta_2^2/4 + g^2}\) when set to zero give the optimal detuning condition: 

\begin{equation}
    \Delta_1 = -\Delta_2/2 \pm \sqrt{\Delta_2^2/4 + g^2}
    \label{optimal_detuning_g>>pump}
\end{equation}

\begin{figure}[h]

    \centering
    \includegraphics[width = 0.5\textwidth]{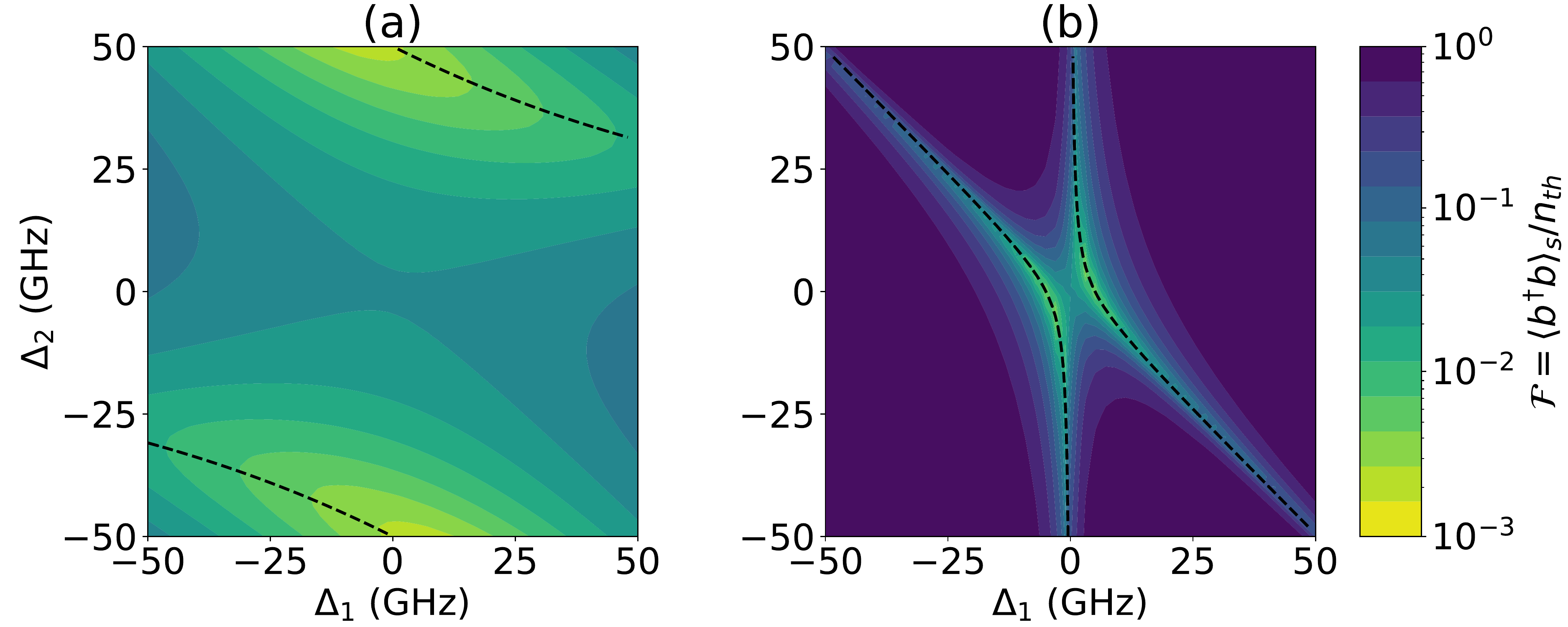}
    \caption{Variation of \(\mathcal{F} = \braket{b^{\dagger}b}/n_{th}\) with detunings \(\Delta_1\) and \(\Delta_2\) for two different regimes. (a) \(\Omega = 10g\). The black dashed curves represent the analytically derived optimal detuning expression as in Eq \ref{optimal_detuning_pump>>g}. (b) \(\Omega = 10^{-1}g\). The black dashed curves represent the analytically derived optimal detuning expressions as in Eq \ref{optimal_detuning_g>>pump}. Here, \(g = 5\) GHz, \(\omega_2 - \omega_1 = 120.9\) GHz, \(\gamma_2 = g/2,\) \(\gamma_1 = 10^{-1}g, \gamma = 10^{-4}g\), an initial temperature of 50 K, and the cutoff for the Fock state basis \(N = 40\).}
    \label{fig:detuning_appendix}
\end{figure}

Using quantum toolbox, we simulate the variation in the figure of merit \(\mathcal{F}\) with \(\Delta_1\) and \(\Delta_2\) in Figure \ref{fig:detuning_appendix}. We set \(\omega_2 - \omega_1 = 120.9\) GHz, \(g = 5\) GHz, \(\gamma_2 = g/2, \gamma_1 = 10^{-1}g, \gamma = 10^{-4}g\) and an initial temperature of \(50\) K. We also plot the analytically derived expressions of optimal detuning as black dashed curves. For \(\Omega \gg g\) case, we set \(\Omega = 10g\) and for \(g \gg \Omega\) case, \(\Omega = 10^{-1} g\). Both these cases show that minimization of \(\mathcal{F}\) is achievable in non-resonant system by tuning the pump frequency which in turn changes \(\Delta_1\). We observe that the minimum achievable value of \(\mathcal{F}\) in the regime \(\Omega \not\approx g\) obtained by varying the detunings is greater as compared to in the regime \(\Omega \approx g\).

\subsection{Decay rates}
\begin{figure}[h]
    \centering
    \includegraphics[width = 0.5\textwidth]{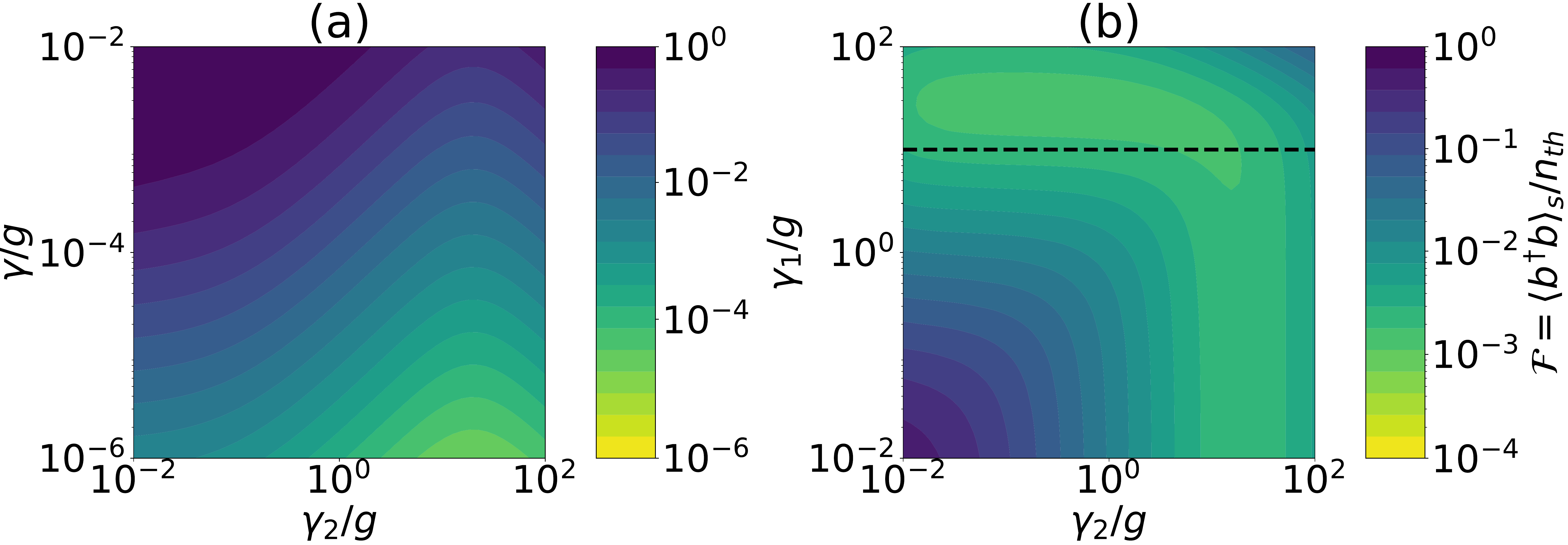}
    \caption{Variation of \(\mathcal{F}\) with (a) phonon decay rate \(\gamma\) and decay rate of the second excited state \(\gamma_2\) for \(\gamma_1 = 10^{-1}g\) (b) decay rate of the first excited state \(\gamma_1\) and the second excited state \(\gamma_2\) for \(\gamma = 10^{-4}g\). The black dashed horizontal line denotes \(\gamma_1 = \Omega\). Here, \(g = 5\) GHz, \(\Delta_1 = \Delta_2 = 0\), \(\omega_2 - \omega_1 = 120.9\) GHz, \(\Omega = 10g\), an initial temperature of 50 K, and the cutoff for the Fock state basis \(N = 40\).}
    \label{variation_decay_rates_pump>>g}
\end{figure}

\begin{figure}[h]
    \centering
    \includegraphics[width = 0.5\textwidth]{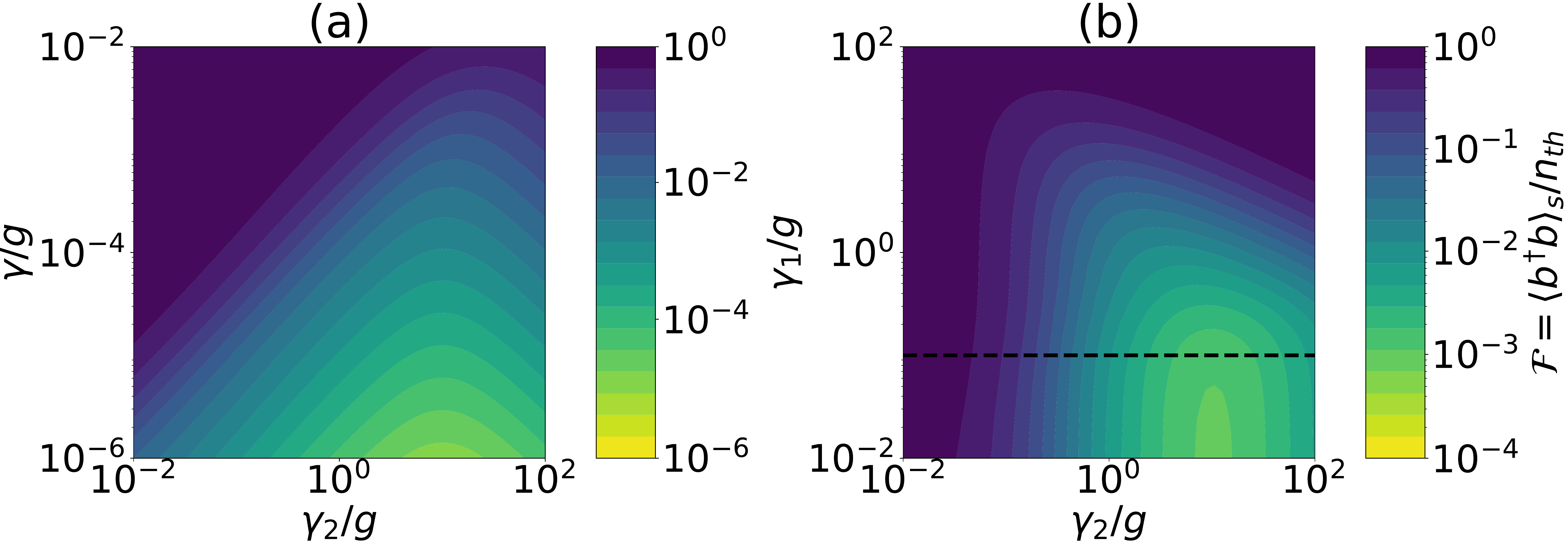}
    \caption{Variation of \(\mathcal{F}\) with (a) phonon decay rate \(\gamma\) and decay rate of the second excited state \(\gamma_2\) for \(\gamma_1 = 10^{-1}g\) (b) decay rate of the first excited state \(\gamma_1\) and the second excited state \(\gamma_2\) for \(\gamma = 10^{-4}g\). The black dashed horizontal line denotes \(\gamma_1 = \Omega\). Here, \(g = 5\) GHz, \(\Delta_1 = \Delta_2 = 0\), \(\omega_2 - \omega_1 = 120.9\) GHz, \(\Omega = 0.1g\), an initial temperature of 50 K, and the cutoff for the Fock state basis \(N = 40\).}
     \label{variation_decay_rates_pump<<g}
\end{figure}

Using the quantum toolbox, we plot the variation of \(\mathcal{F}\) with decay rates for the regime \(\Omega \gg g\) in Figure \ref{variation_decay_rates_pump>>g} and \(g \gg \Omega\) in Figure \ref{variation_decay_rates_pump<<g}. Here, \(\Delta_1 = \Delta_2 = 0\). Since the optimal detuning condition is modified in these regimes from \(\Delta_1 = \Delta_2 = 0\) (as discussed in previous subsection), the minimum achievable \(\mathcal{F}\) is greater than in the regime \(\Omega \approx g\).

In both the regimes \(\Omega \gg g\) and \(g \gg \Omega\), the decay rates \(\gamma\) and \(\gamma_2\) follow the same trend as in Figure \ref{decay_rates_plot} of the main text. However, in Figure \ref{variation_decay_rates_pump>>g}(b), we observe that an optimal \(\gamma_1\) exists in the regime \(\Omega \gg g\). This can be understood as follows: 

\begin{itemize}
    \item[--] A larger value of \(\Omega\) allows for a large \(\gamma_1\) before decoherence sets in.
    \item[--] As discussed in the previous subsection, the optimal detuning condition in the regime \(\Omega \gg g\), which results in resonant interactions, is modified from the current detuning values (\(\Delta_1 = \Delta_2 = 0\)). To accommodate for the now non-resonant interaction between the eigenstates, the eigenstates require a large linewidth. The linewidth for the eigenstates is proportional to the decay rates \(\gamma_1\) and \(\gamma_2\). Therefore, under the condition \(\Omega \gtrsim \gamma_1\), \(\gamma_1\) can be varied to minimize \(\mathcal{F}\), leading to an optimal \(\gamma_1\).
\end{itemize}

A similar trend is not observed in Figure \ref{variation_decay_rates_pump<<g}(b) since \(g \gg \Omega\), which does not allow room for the tuning of \(\gamma_1\) to minimize \(\mathcal{F}\) without setting in decoherence.

\section{\label{Appendix:RateEq} Equations of motion for density matrix elements}
\begin{figure}[h]

    \centering
    \includegraphics[width = 0.4\textwidth]{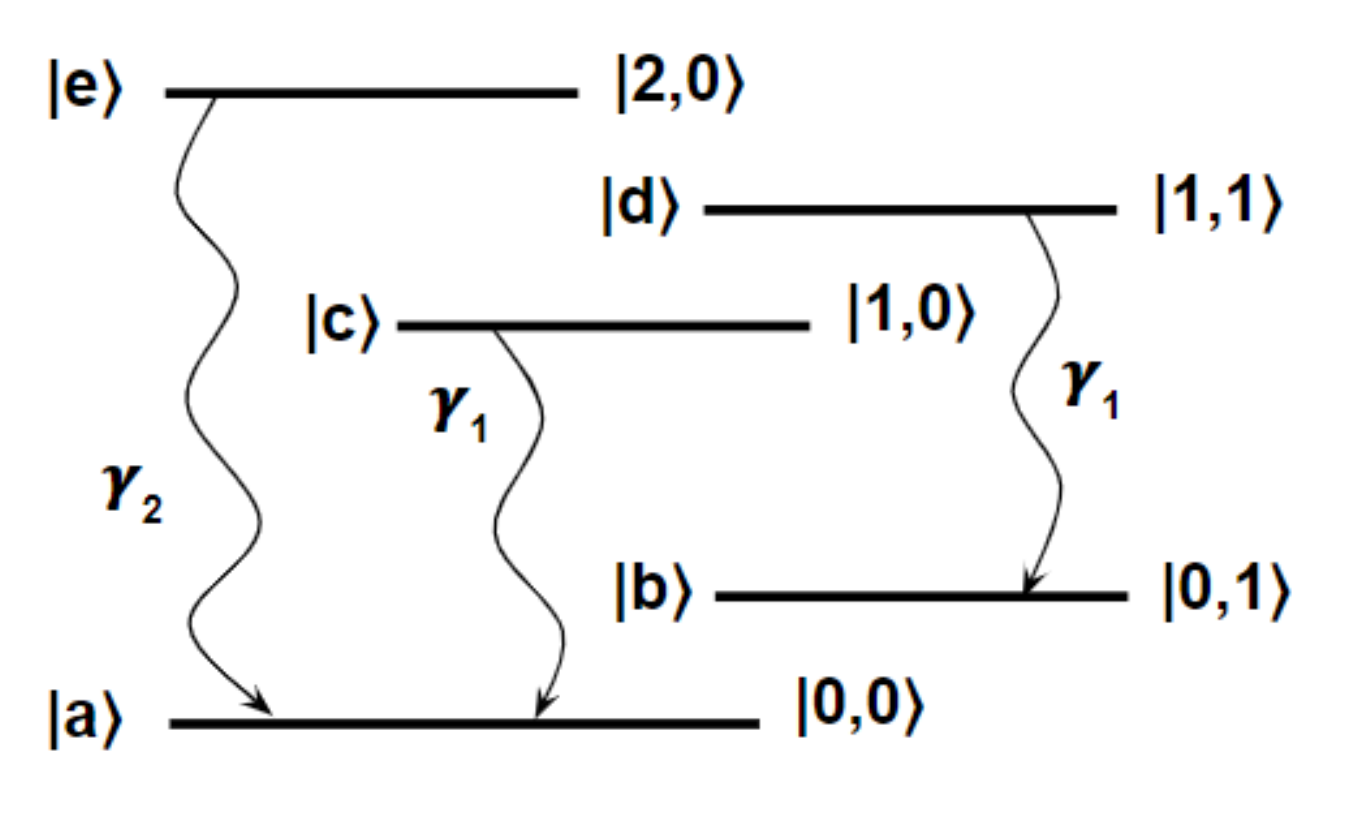}
    \caption{Combined states for the quantum dot system and mechanical resonator mode of the form \(\ket{i,j}\) where index \(i\) and \(j\) corresponds to the quantum dot system and the mechanical resonator mode respectively.}
    \label{combined states}
\end{figure}

We denote the combined states of the quantum dot system and the mechanical resonator mode by \(\ket{i,j}\) where indices \(i\in [0,1,2] \) and \(j\in [0,1] \) correspond to the quantum dot system and the mechanical resonator, respectively (Figure \ref{combined states}). For brevity, we label the combined state basis as \(\ket{a}, \ket{b}, .., \ket{e}\) as depicted in Figure \ref{combined states} by the labels on the left.

Next, we derive the equations of motion of various density matrix elements needed to characterize the dynamics of the combined system completely:

\begin{equation}
    \begin{aligned}
    \label{rate eqs}
        \dfrac{d\rho_{bb}}{dt} &= i \Omega(\rho_{bd} - \rho_{db}) + \gamma n_{th} \rho_{aa} - \gamma (n_{th} + 1)\rho_{bb} + \gamma_1 \rho_{dd}\\
        \dfrac{d\rho_{cc}}{dt} &= i \Omega(\rho_{ca} - \rho_{ac}) + \gamma (n_{th} + 1)\rho_{dd} - \gamma n_{th} \rho_{cc} - \gamma_1 \rho_{cc} \\
        \dfrac{d\rho_{dd}}{dt} &= i g (\rho_{de} - \rho_{ed}) + i \Omega (\rho_{db} - \rho_{bd}) + \gamma n_{th} \rho_{cc} - \gamma (n_{th} + 1)\rho_{dd}\\
        &- \gamma_1 \rho_{dd} \\
        \dfrac{d\rho_{ee}}{dt} &= i g (\rho_{ed} - \rho_{de}) -\gamma_2 \rho_{ee} \\
        \dfrac{d\rho_{db}}{dt} &= - i g \rho_{eb} + i \Omega(\rho_{dd} - \rho_{bb}) + \gamma n_{th} \rho_{ca} - \gamma (n_{th} + 1)\rho_{db} \\
        & - \dfrac{\gamma_1}{2} \rho_{db} \\
        \dfrac{d\rho_{eb}}{dt} &= - i g \rho_{db} + i \Omega \rho_{ed} - \dfrac{\gamma_2 + \gamma(n_{th} + 1)}{2} \rho_{eb} \\
        \dfrac{d\rho_{ca}}{dt} &= i \Omega (\rho_{cc} - \rho_{aa}) + \gamma (n_{th} + 1)\rho_{db} - \dfrac{\gamma_1}{2} \rho_{ca} - \gamma n_{th} \rho_{ca} \\
        \dfrac{d\rho_{ed}}{dt} &= i g (\rho_{ee} - \rho_{dd}) + i \Omega \rho_{eb} - \dfrac{\gamma_1 + \gamma_2 + \gamma(n_{th} + 1)}{2} \rho_{ed}\\
    \end{aligned}
\end{equation}

We calculate the expression for the steady state phonon number \(\braket{b^{\dagger}b}_s = \rho_{bb} + \rho_{dd}\) under the following two conditions:

\begin{widetext}
\subsection{\label{Appendix:steadystate_phonon_gamma1>>gamma2} \(\gamma_1 \ll \gamma_2\)}

\begin{equation}
\label{eq:phonon_gamma1>>gamma2}
\begin{aligned}
    \braket{b^{\dagger}b}_s &= \frac{\gamma n_{\text{th}} \left(2 \gamma _2 \left(\gamma _2^2 \Omega ^2+g^4+4 \Omega ^4\right)+\gamma  \left(\gamma _2^2 \left(\left(g^2+4 \Omega ^2\right) n_{\text{th}}+2 \left(g^2+2 \Omega ^2\right)\right)+4 g^2 \left(g^2+2 \Omega ^2\right) \left(n_{\text{th}}+1\right)\right)\right)}{2 \gamma _2 \left(2 \gamma _2 g^2 \Omega ^2+\gamma  \left(\left(3 g^4+4 g^2 \Omega ^2+8 \Omega ^4\right) n_{\text{th}}+2 \left(g^4+g^2 \Omega ^2+2 \Omega ^4\right)+\gamma _2^2 \Omega ^2 \left(2 n_{\text{th}}+1\right)\right)\right)}
\end{aligned}
\end{equation}

\subsection{\label{Appendix:steadystate_phonon_gamma1=gamma2}\(\gamma_1 = \gamma_2\)}

\begin{equation}
\label{eq:phonon_gamma1=gamma2}
\begin{aligned}
    \braket{b^{\dagger}b} &= \frac{x}{y} \\
    x &= \gamma _m (2 \gamma _2 (\gamma _2^6+3 g^4 \Omega ^2+\gamma _2^4 (2 g^2+9 \Omega ^2)+\gamma _2^2 (g^4+2 g^2 \Omega ^2+24 \Omega ^4)+16 \Omega ^6)+\gamma _m (2 \gamma _2^4 ((11 g^2+42 \Omega ^2) n_{\text{th}} \\
    & +5 g^2+27 \Omega ^2)+4 g^2 \Omega ^2 (g^2+4 \Omega ^2) (n_{\text{th}}+1)+\gamma _2^2 (g^4+30 g^2 \Omega ^2+(7 g^4+16 g^2 \Omega ^2+96 \Omega ^4) n_{\text{th}}+72 \Omega ^4)\\
    & +3 \gamma _2^6 (5 n_{\text{th}}+3)))\\
    y &= 2 \gamma _2 (2 \gamma _2 g^2 \Omega ^2 (\gamma _2^2+4 \Omega ^2)+\gamma _m ((\gamma _2^2+4 \Omega ^2) (\gamma _2^4+g^4+\gamma _2^2 (2 g^2+5 \Omega ^2)+g^2 \Omega ^2+4 \Omega ^4) +n_{\text{th}} (7 g^4 \Omega ^2\\
    &+12 g^2 \Omega ^4 +\gamma _2^2 (2 g^4+2 \gamma _2^2 (\gamma _2^2+2 g^2+9 \Omega ^2)+19 g^2 \Omega ^2+48 \Omega ^4)+32 \Omega ^6)))
\end{aligned}
\end{equation}
\end{widetext}

\section{\label{dephasing QD}Effect of pure dephasing in the regime \(\gamma_1 \ll \gamma_2\)}

\begin{figure*}[hbt!]
    
    \centering
    \includegraphics[width = \textwidth]{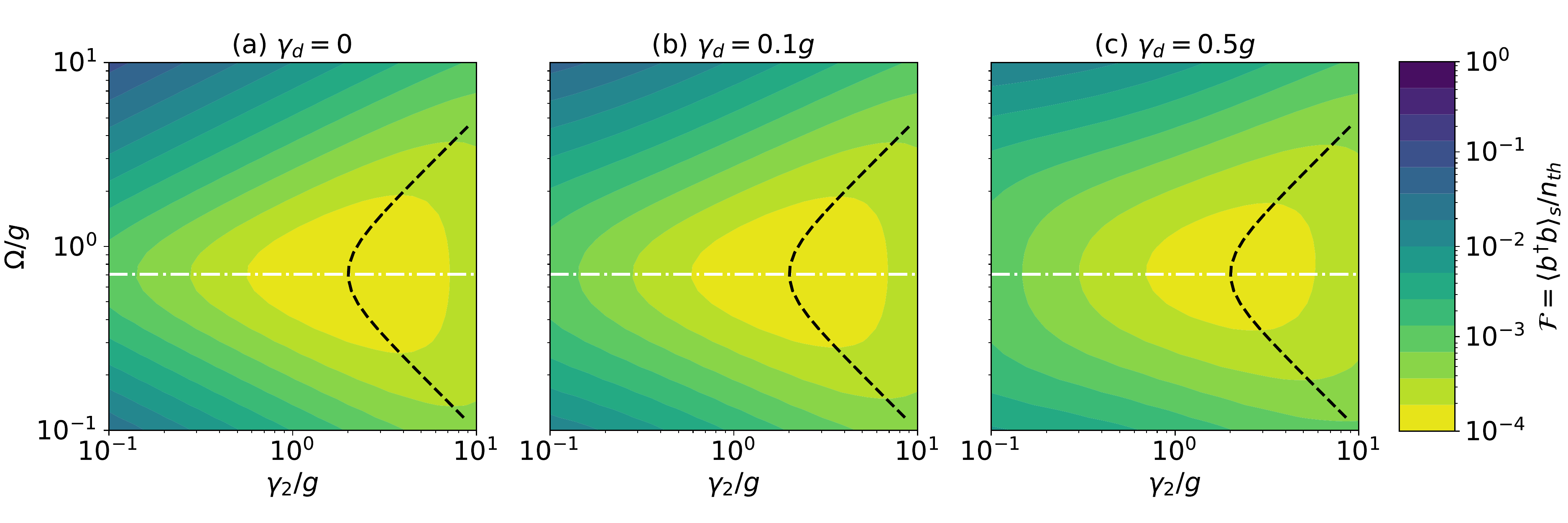}
    \caption{\textbf{Regime \(\gamma_1 \ll \gamma_2\)}: Variation of \(\mathcal{F}\) with \(\Omega\) and \(\gamma_2\) for different pure dephasing rates \(\gamma_d\). The white dot-dashed line and the black dashed curve represents the optimal parameters \(\Omega_o\) and \(\gamma_{2_o}\) respectively as derived in Eq 8 in the main text. Other system parameters are: \(\omega_m = 241.8\) GHz, \(g = 20\) GHz, \(\gamma_1 = 10^{-6}\) GHz, \(\gamma = 10^{-3}\) GHz, an initial temperature of \(17\) K, and the cutoff for the Fock state basis \(N = 40\).}
    \label{dephasing_QD_plot}
\end{figure*}

Here, we try to discern the effect of pure dephasing on \(\mathcal{F}\) in the regime \(\gamma_1 \ll \gamma_2\) using the quantum toolbox. To do so, we modify the master equation (Eq. 3) to include pure dephasing:
\begin{equation}
    \begin{aligned}
        \dfrac{d \rho}{dt} &= i [\rho, \mathbf{H_{\text{rotated}}}] + \gamma_1 \mathcal{L}[\sigma_{01}] \rho + \gamma_2 \mathcal{L}[\sigma_{02}] \rho \\
        & + \gamma (n_{th} + 1) \mathcal{L}[b] \rho + \gamma n_{th} \mathcal{L}[b^{\dagger}] \rho \\
        &+ \underbrace{\gamma_d \mathcal{L}[\sigma_{11}] \rho + \gamma_d \mathcal{L}[\sigma_{22}] \rho}_{\text{Pure  Dephasing}}
    \end{aligned}
\label{master_eq_pure_dephasing}
\end{equation}

The last two terms account for the pure dephasing process in the excited states \(\ket{1}\) and \(\ket{2}\) respectively. The Hamiltonian for the system remains the same (Eq. 2). In Figure \ref{dephasing_QD_plot}, we plot the variation of \(\mathcal{F}\) with \(\Omega\) and \(\gamma_2\), and for different values of pure dephasing rate \(\gamma_d\) using Eq 2 and Eq \ref{master_eq_pure_dephasing}. The white dot-dashed line and the black dashed curve represent the optimal parameters \(\Omega_o\) and \(\gamma_{2_o}\), respectively, as derived in Eq 8 in the main text. For cadmium selenide colloidal quantum dots, the pure dephasing rate \(\gamma_d \approx 10\) GHz \(= g/2\) \cite{Khosla_2018}. We observe that even with the inclusion pure dephasing rate, the minimum achievable value for \(\mathcal{F}\) and the optimal values for the parameters \(\Omega\) and \(\gamma_2\) do not change significantly.

\section{\label{hamiltonian equivalence} Hamiltonian of a TLS coupled to an optical cavity mode and a mechanical resonator}

The Hamiltonian for a single mode cavity mode strongly coupled to a two-level system (TLS), forming polaritons, that in turn couple to a mode of a mechanical resonator is given by:

\begin{equation}
\label{eq:H_polariton}
    \begin{aligned}
        \mathbf{H_{\text{system}}} &= \underbrace{\omega_{c}  a^{\dagger} a +  \omega_{\alpha} \sigma_{\alpha \alpha} + G (\sigma_{\beta \alpha} a^{\dagger} + \sigma_{\alpha \beta}a)}_{\mathbf{H_{\text{JC}}}} \\
        &+ \Omega(\sigma_{\alpha \beta} e^{-i \omega_p t} + \sigma_{\beta \alpha} e^{i \omega_p t}) + \omega_{m}b^{\dagger}b + ga^{\dagger} a (b^{\dagger} + b)
    \end{aligned}
\end{equation}
where \(\ket{\beta}\) and \(\ket{\alpha}\) are the ground and excited states for the two-level system with frequency separation \(\omega_{\alpha}\), \(a\) \((b)\) is the annihilation operator the the cavity (mechanical) mode with frequency \(\omega_a (\omega_m)\), \(G\) is TLS-cavity mode coupling strength, \(\Omega\) is the pumping strength to the TLS at frequency \(\omega_p\) and \(g\) is optomechanical coupling strength. Following the process as described in references \cite{Restrepo_2014, Restrepo_2017}, we write the system Hamiltonian in the diagonalized basis of the TLS-cavity Hamiltonian (\(\mathbf{H_{\text{JC}}}\)). Restricting the number of photons in the cavity mode to one, the Hamiltonian \(\mathbf{H_{\text{JC}}}\) can be readily diagonalized such that \(\mathbf{H_{\text{JC}}} \ket{\pm} = \omega_{\pm} \ket{\pm}\) where

\begin{subequations}
\begin{eqnarray}
    \ket{0} = \ket{\beta, 0} \\
    \ket{-} = - \sin \theta \ket{\alpha,0} + \cos \theta \ket{\beta,1} \\
    \ket{+} = \cos\theta \ket{\alpha,0} + \sin \theta \ket{\beta,1} \\
    \omega_{\pm} = \dfrac{\omega_a + \omega_c}{2} \pm \sqrt{G^2 + \dfrac{\Delta^2}{4}}
\end{eqnarray}
\end{subequations}
Here \(\tan(2\theta) = 2G/\Delta\) and \(\Delta = \omega_{\alpha} - \omega_a\). The cavity annihilation operator in the restricted polariton basis can be written as \(a = \sin{\theta} \ketbra{\beta, 0}{+} + \cos{\theta} \ketbra{\beta, 0}{-}\) \cite{Neuman:18} and the number operator as \(a^{\dagger} a = \sin^2{\theta} \ketbra{+}{+} + \cos^2{\theta} \ketbra{-}{-} + \sin{\theta} \cos{\theta} (\ketbra{+}{-} + \ketbra{-}{+})\). Substituting these relations in Eq. \ref{eq:H_polariton}, the system Hamiltonian in the polariton basis becomes:

\begin{widetext}
\begin{equation}
\label{eq:full_polariton}
    \begin{aligned}
        \mathbf{H_{\text{system}}} &= \omega_{-}\sigma_{--} + \omega_{+} \sigma_{++} + \omega_m b^{\dagger} b
        + \Omega \Big[ \big(\cos \theta \sigma_{0+} - \sin \theta \sigma_{0-} + \cos \theta \sin \theta \sigma_{++} + \cos^2\theta\sigma_{-+} - \sin^2 \theta \sigma_{+-} \\
        &- \cos\theta\sin\theta\sigma_{--} \big)e^{i \omega_p t} + c.c \Big] + g(b + b^{\dagger}) \big[\sin^2\theta \sigma_{++} + \cos^2\theta\sigma_{--} + \cos\theta\sin\theta(\sigma_{+-} + \sigma_{-+}) \big]
    \end{aligned}
\end{equation}

We assume that the applied pump is close to the frequency of the lower polariton (\(\omega_p - \omega_- \approx 0\)) and that the mechanical resonator frequency is close to the frequency difference between the upper and lower polaritons (\(\omega_+ - \omega_- - \omega_m \approx 0\)). Because of this, several terms in Eq \ref{eq:full_polariton} are off resonant and can be dropped under the rotating wave approximation. To bring this fact out clearly, we move to an interaction picture by rotating the Hamiltonian in Eq \ref{eq:full_polariton} with respect to the Hamiltonian \(H_o = \omega_{-}\sigma_{--} + \omega_{+} \sigma_{++} + \omega_m b^{\dagger} b\) and obtain:

\begin{equation}
    \label{eq:full_polariton_rotated}
    \begin{aligned}
        \mathbf{H_{\text{rotated}}} &= \Omega \Big[ \big( \cos \theta \sigma_{0+} e^{i(\omega_p - \omega_+)t} - \sin \theta \sigma_{0-}e^{i(\omega_p - \omega_-)t} + \cos \theta \sin \theta \sigma_{++}e^{i\omega_pt} + \cos^2\theta\sigma_{-+}e^{i(\omega_- - \omega_+ + \omega_p)t} \\
        &- \sin^2 \theta \sigma_{+-}e^{i(\omega_+ - \omega_- + \omega_p)t} - \cos\theta\sin\theta\sigma_{--}e^{i\omega_pt} \big) + c.c \Big] + g \big(be^{-i\omega_mt} + b^{\dagger}e^{i\omega_mt} \big) \big[\sin^2\theta \sigma_{++} + \cos^2\theta\sigma_{--}\\
        & + \cos\theta\sin\theta \big( \sigma_{+-}e^{i(\omega_+ - \omega_-)t} + \sigma_{-+}e^{i(\omega_- - \omega_+)t} \big) \big]
    \end{aligned}
\end{equation}

\noindent In the above equation, we keep only the slow rotating terms and drop the fast rotating terms under the rotating wave approximation. Moving back to the non-rotating frame, the combined system Hamiltonian now becomes:

\begin{equation}
    \label{eq:polariton_rotated}
    \begin{aligned}
        \mathbf{H_{system}} &= \omega_{-}\sigma_{--} + \omega_{+} \sigma_{++} + \omega_m b^{\dagger} b - \Omega \sin{\theta} \big( \sigma_{0-}e^{i\omega_pt} + \sigma_{-0}e^{-i\omega_pt} \big) + g \sin \theta \cos \theta (\sigma_{-+} b + \sigma_{+-} b^{\dagger})
    \end{aligned}
\end{equation}
\end{widetext}

We now return to our assumption of restricting the cavity photons to one. The cavity-TLS coupling strength \(G \gg g, \Omega\) and thus the polariton dynamics is much faster than the mechanical counterpart. This leads to sparsely populated higher energy polariton rungs. For the values in the manuscript, \(> 99 \%\) of the cavity-TLS population resides in the states \(\{\ket{0}, \ket{-}, \ket{+} \}\), which justifies our assumption.

% In the bad cavity regime, the decay rate of the cavity mode (\(\kappa\)) will be much greater than the decay rate of the TLS. Therefore, the decay rates of the polariton states take the form \cite{Neuman:18}:

% \begin{equation}
% \label{eq:gamma2kappa}
%     \begin{aligned}
%         \gamma_{+} &\approx \kappa \cos^2(\theta)\\
%         \gamma_{-} &\approx \kappa \sin^2(\theta)
%     \end{aligned}
% \end{equation}

\section{\label{dephasing in polariton} Effect of incoherent processes in the regime \(\gamma_1 = \gamma_2\)}

\begin{figure*}[hbt!]
    
    \centering
    \includegraphics[width = \textwidth]{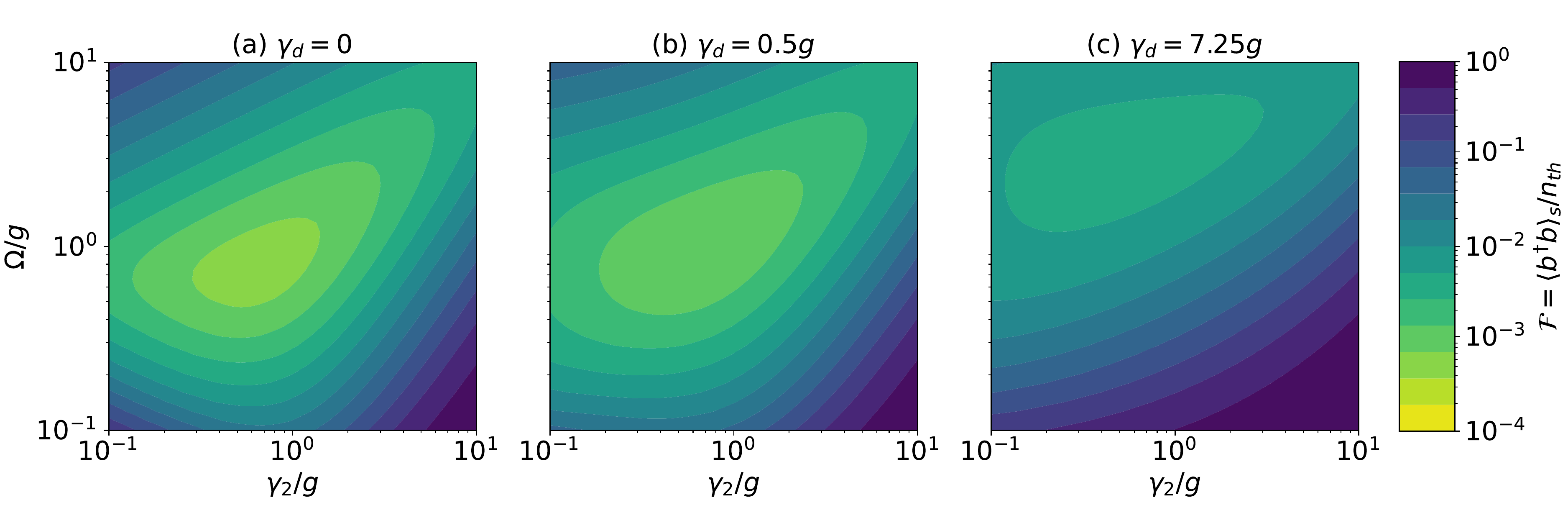}
    \caption{\textbf{Regime \(\gamma_1 = \gamma_2\)}: Variation of \(\mathcal{F}\) with \(\Omega\) and \(\gamma_2\) for different pure dephasing rates \(\gamma_d\). Other system parameters are: \(G = 5\) GHz, \(g = 0.002 G, \omega_m = 2 G, \gamma =  \omega_m/Q_m = 10^{-7}\) GHz, an initial temperature of \(2.63\) K corresponding to \(n_{th} = 5\), and the cutoff for the Fock state basis \(N = 40\).}
    \label{dephasing_polariton_plot}
\end{figure*}

\subsection{Pure dephasing}
To account for pure dephasing, we use the same formalism as in Appendix \ref{dephasing QD}. The dephasing rate for InGaAs quantum dots is reported to be \(\gamma_d = 0.3 \mu\)eV \(= 72.54\) MHz \(= 7.25g\) \cite{HighCooperativity}. Using the quantum toolbox, we plot the variation of \(\mathcal{F}\) with \(\Omega\) and \(\gamma_2\), and for different values of pure dephasing rate \(\gamma_d\) in Figure \ref{dephasing_polariton_plot}. The effect of pure dephasing is to increase the optimal value of \(\Omega\) and decrease the optimal value of \(\gamma_2\) to account for the additional decoherence. This effect is significant only when the dephasing rate \(\gamma_d\) is of the order of the optomechanical coupling strength \(g\).

\subsection{Above band pumping}
To account for above band or incoherent pumping, we modify the original master equation to:
\begin{equation}
    \begin{aligned}
        \dfrac{d \rho}{dt} &= i [\rho, \mathbf{H_{\text{rot}}}] + \gamma_1 \mathcal{L}[\sigma_{01}] \rho + \gamma_2 \mathcal{L}[\sigma_{02}] \rho \\
        & + \gamma (n_{th} + 1) \mathcal{L}[b] \rho + \gamma n_{th} \mathcal{L}[b^{\dagger}] \rho \\
        & + \underbrace{\gamma_{p} \mathcal{L}[\sigma_{10}] \rho + \gamma_{p} \mathcal{L}[\sigma_{20}] \rho}_{\text{Incoherent Pumping}}
    \end{aligned}
\label{master_eq_incoherent_pumping}
\end{equation}

\begin{figure}[hbt!]
    
    \includegraphics[width=0.5\textwidth]{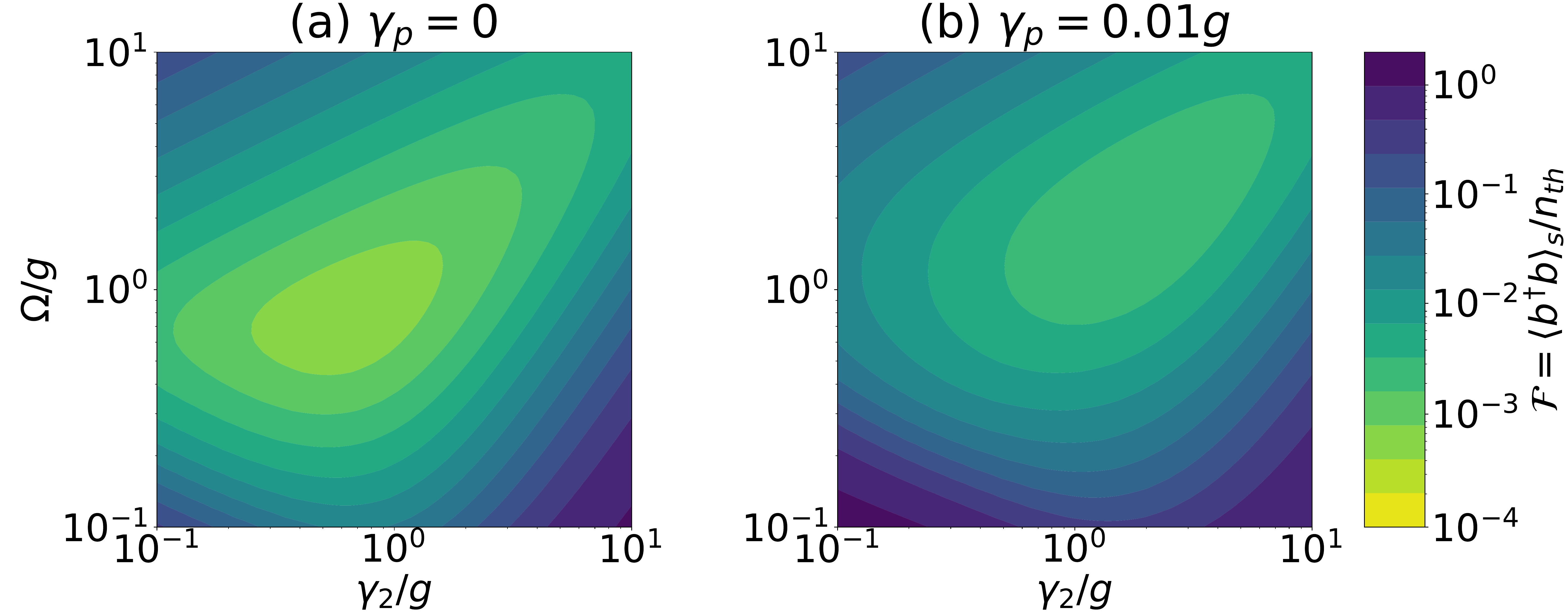}
    \caption{\textbf{Regime \(\gamma_1 = \gamma_2\)}: Variation of \(\mathcal{F}\) with \(\Omega\) and \(\gamma_2\) for (a) \(\gamma_{p} = 0\) and (b) \(\gamma_{p} = 0.01g\). Other system parameters are: \(G = 5\) GHz, \(g = 0.002 G, \omega_m = 2 G, \gamma =  \omega_m/Q_m = 10^{-7}\) GHz, an initial temperature of \(2.63\) K corresponding to \(n_{th} = 5\), and the cutoff for the Fock state basis \(N = 40\).}
    \label{incoherent_polariton_plot}
\end{figure}

\noindent The last two terms in the above equation (Eq \ref{master_eq_incoherent_pumping}) represent above band pumping to the excited states \(\ket{1}\) and \(\ket{2}\) respectively. The Hamiltonian for the system remains the same (Eq 2). We plot the variation of \(\mathcal{F}\) with \(\Omega\) and \(\gamma_2\) using Eq 2 and Eq \ref{master_eq_incoherent_pumping} in Figure \ref{incoherent_polariton_plot} for \(\gamma_p = 0\) in (a) and \(\gamma_p = 0.01\)g in (b). Comparing Figure \ref{incoherent_polariton_plot}(a) and Figure \ref{incoherent_polariton_plot}(b), we observe that the above band pumping leads to an increase in the minimum value of \(\mathcal{F}\) and heating (\(\mathcal{F} > 1\)) for small \(\Omega/g\) and \(\gamma_2/g\). Furthermore, the optimal values for the parameters \(\Omega\) and \(\gamma_2\) increase. To explain this, we solve the Heisenberg operator equations, including above band pumping, to write the steady state phonon occupation:

\begin{equation}
\label{steadystate_phonon_incoherent_pumping}
    \braket{b^{\dagger}b}_s =  n_{th} - \frac{\gamma_2}{\gamma} \braket{\sigma_{22}}_s + \frac{\gamma_p}{\gamma} \braket{\sigma_{00}}_s
\end{equation}

Compared to Eq 5 of the main text, Eq \ref{steadystate_phonon_incoherent_pumping} has an additional term that is proportional to the above band pumping rate \(\gamma_p\). This additional term is responsible for increase in the steady state phonon number \(\braket{b^{\dagger}b}_s\) and therefore \(\mathcal{F} = \braket{b^{\dagger}b}_s/n_{th}\). The minimization problem for \(\mathcal{F}\) is modified resulting in different optimal parameter values for \(\Omega\) and \(\gamma_2\). To minimize \(\mathcal{F}\), the last two terms in Eq \ref{steadystate_phonon_incoherent_pumping} need to be maximized and minimized simultaneously. Since the above band pumping rate \(\gamma_p\) is non-zero and fixed, the optimal value for \(\Omega\) needs to be increased to decrease \(\braket{\sigma_{00}}_s\) in order to minimize the last term. Subsequently, a larger \(\gamma_2\) is required to accommodate the increased population transitioning to the state \(\ket{2}\). This results in increased optimal values for \(\Omega\) and \(\gamma_2\) as compared to when the above band pumping is absent (\(\gamma_p = 0\)).

\section{Cooling of a mechanical mode coupled to a Manganese doped quantum dot}

\begin{figure}[h]
    
    \includegraphics[width=0.5\textwidth]{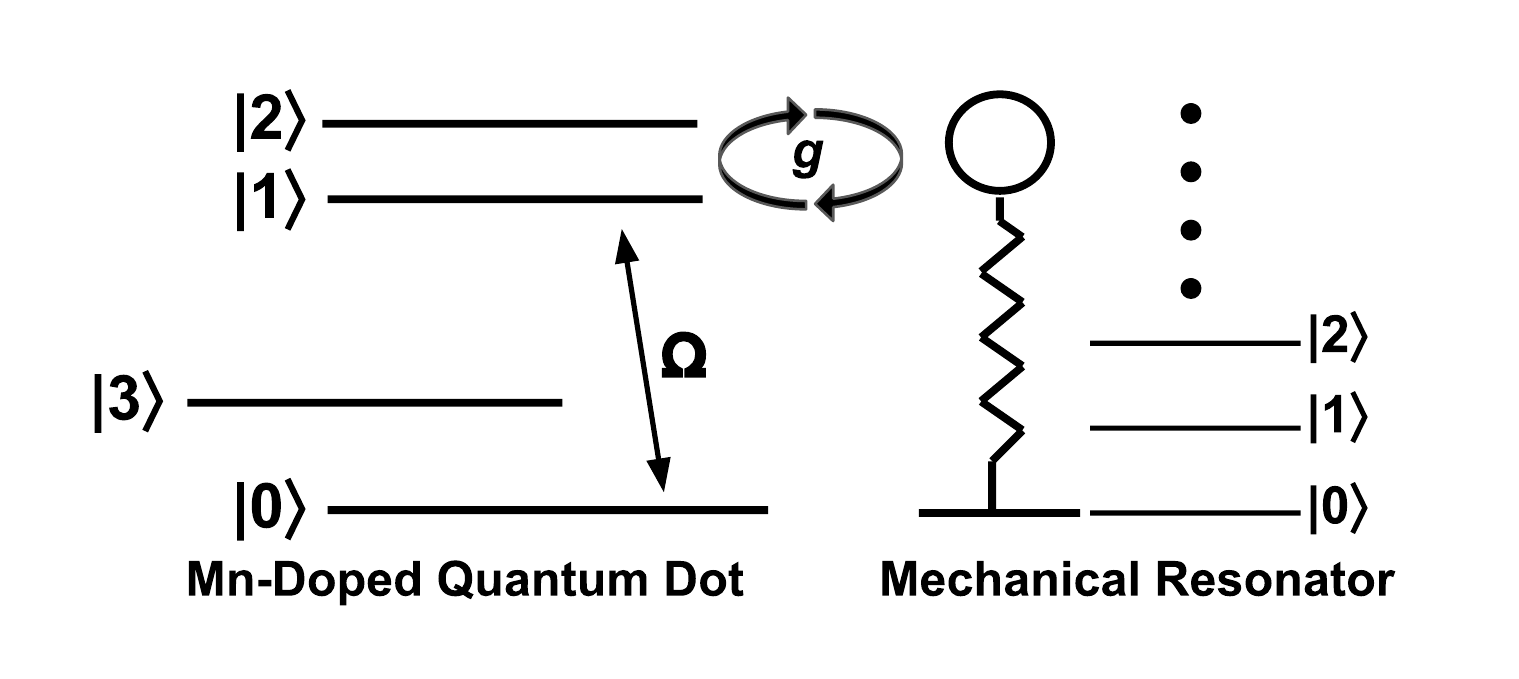}
    \caption{Schematic representing Manganese doped quantum dot coupled to a mode of a mechanical resonator.}
    \label{four-level-system}
\end{figure}

Similar to Section II of the main text, we label the states \(\ket{0} - \ket{3}\) as shown in Figure  \ref{four-level-system}. Moving to a suitable rotated frame of reference, the system Hamiltonian can be simplified to:

\begin{equation}
\begin{aligned}
    \mathbf{H_{\text{rotated}}} &=  \Delta_1 \sigma_{11} + (\Delta_1 + \Delta_2) \sigma_{22} + \omega_3 \sigma_{33} + \Omega (\sigma_{01} + \sigma_{10}) \\
    & + g (\sigma_{12} b^{\dagger} + \sigma_{21}b) \\
\end{aligned}
\label{mn_doped_qd}
\end{equation}
Here \(\Delta_1 = \omega_1 - \omega_p\), \(\Delta_2 = \omega_2 - \omega_1 - \omega_m\), \(g\) is the coupling strength of the two excited states with the mechanical mode of frequency \(\omega_m\), and \(\Omega\) is the coherent pumping strength between the states \(\ket{0}\) and \(\ket{1}\) at frequency \(\omega_p\). The excited states \(\ket{1}\) and \(\ket{2}\) decay to both the ground states \(\ket{0}\) and \(\ket{3}\) with equal decay rate which we denote by \(\gamma_2\). Furthermore, the higher energy ground state \(\ket{3}\) also decays to the lower energy ground state \(\ket{0}\) with rate \(\gamma_3\). The master equation for the system, taking into account all the incoherent process is:

\begin{equation}
\label{master_eq_mn_doped_qd}
    \begin{aligned}
        \dfrac{d \rho}{dt} &= i [\rho, \mathbf{H_{\text{rotated}}}] + \gamma_2 \Big( \mathcal{L}[\sigma_{01}] \rho + \mathcal{L}[\sigma_{02}] \rho + \mathcal{L}[\sigma_{31}] \rho \\
        & + \mathcal{L}[\sigma_{32}] \rho \Big) + \gamma_3 \mathcal{L}[\sigma_{03}] \rho + \gamma (n_{th} + 1) \mathcal{L}[b] \rho \\
        & + \gamma n_{th} \mathcal{L}[b^{\dagger}] \rho + \gamma_d \mathcal{L}[\sigma_{11}] \rho + \gamma_d \mathcal{L}[\sigma_{22}]
    \end{aligned}
\end{equation}

\begin{figure}[hbt!]
    \includegraphics[width=0.5\textwidth]{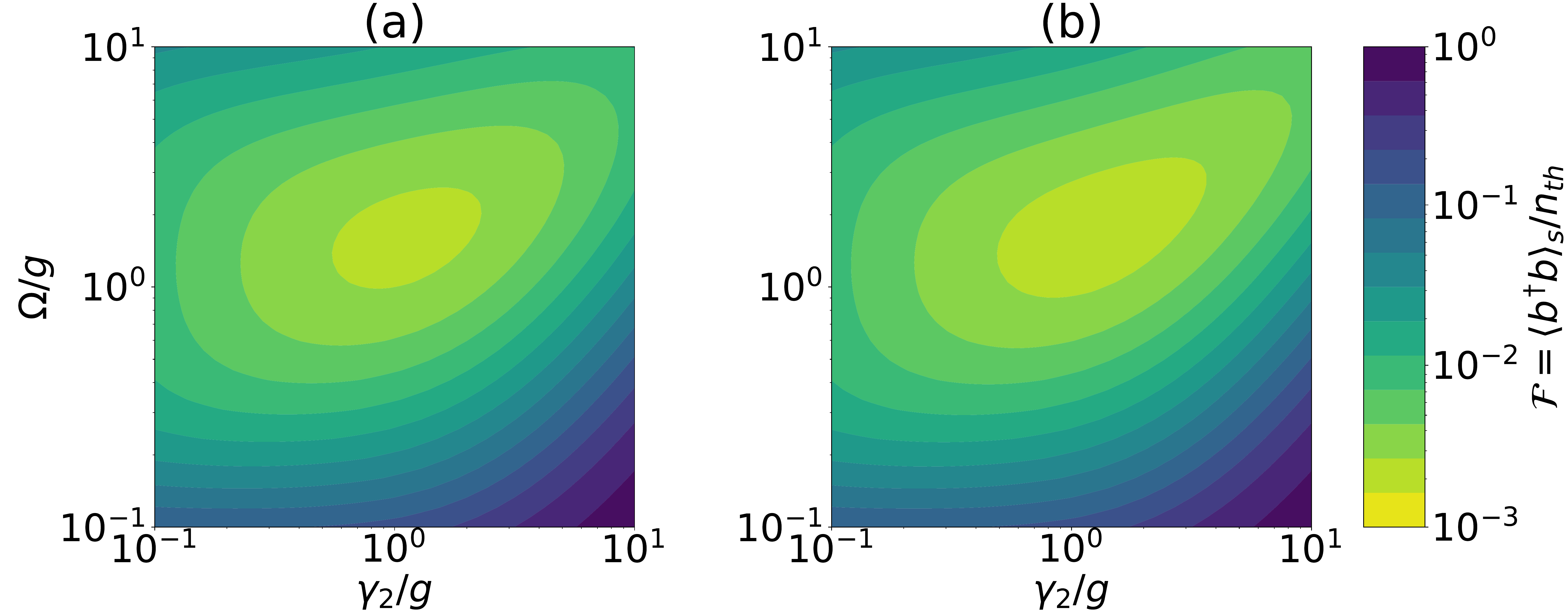}
    \caption{Variation of \(\mathcal{F}\) with \(\Omega\) and \(\gamma_2\) as per (a) the four-level Mn doped quantum dot model and (b) \(\gamma_1 = \gamma_2\) regime formulation described in Section III B of the main text. The system parameters are: \(\omega_3 = 170\) GHz, \(g = 10\) MHz, \(\omega_m = 35\) GHz, \(\gamma_3 = \gamma_d = 24.18\) MHz, \(Q_m = 10^7\), an initial temperature of \(5\) K corresponding to \(n_{th} = 2.52\), and the cutoff for the Fock states \(N = 40\).}
    \label{Mn_doped_QD}
\end{figure}

The last two terms in Eq \ref{master_eq_mn_doped_qd} account for the pure dephasing in the excited states. Consistent with the reference \cite{MnDopedQD}, we set \(\omega_3 = 170\) GHz, \(\omega_m = 35\) GHz, \(\gamma_3 = \gamma_d = 24.18\) MHz. The system affords decay and dephasing rates of the order of 10 MHz because of the highly delocalized nature of the Mn dopant complex in the quantum dot. We further set \(\Delta_1 = \Delta_2 = 0\), \(g = 10\) MHz, and the mechanical quality factor \(Q_m = \omega_m/\gamma_m = 10^7\). Similar to our previous formulations, the control parameters in this model are the pumping strength \(\Omega\) and the decay rate of the excited states \(\gamma_2\). The decay rate of the excited states can be increased (decreased) simultaneously via Purcell enhancement (suppression) by coupling to two different modes of the same optical cavity. Using the quantum toolbox, we solve the master equation (Eq \ref{master_eq_mn_doped_qd}) and calculate \(\mathcal{F}\). In Figure \ref{Mn_doped_QD}(a) we plot the variation of \(\mathcal{F}\) with \(\Omega\) and \(\gamma_2\) for the full four-level system model. In comparison, using the same parameters, we plot \(\mathcal{F}\) in the limit \(\gamma_3 \to \infty\), thereby reducing the four-level system to a three-level system, in Figure \ref{Mn_doped_QD}(b). This is equivalent to the formulation described in Section III B of the main text (regime \(\gamma_1 = \gamma_2\)).

We observe that both models are in good agreement except when \(\Omega, \gamma_2 > g\). This can be understood as follows: the finite steady state population of the state \(\ket{3}\) that is not involved in the cooling dynamics leads to a disagreement between the two models. For the figure of merit \(\mathcal{F}\) to be equivalent in both the models, the steady state population of the higher energy ground state \(\braket{\sigma_{33}}_s\) should be as small as possible. In the limit \(\gamma_3 \to \infty\), \(\braket{\sigma_{33}}_s \to 0\) and the four-level system exactly mimics a three-level system. As the value of \(\gamma_3\) is finite and fixed, other system parameters need to be taken into account to determine the validity of the approximation. In the regime \(\Omega > g\) and \(\gamma_2 > g\), the first condition ensures that the state \(\ket{0}\) is sparsely populated thus pumping the population to the first excited state \(\ket{1}\). The second condition leads to an increased population of the state \(\ket{3}\) because of a strong decay from the excited states. When both of these parameter regimes act together, it leads to a disagreement between the two models as can be seen in the top right corners of Figure \ref{Mn_doped_QD}(a) and (b). In all other ranges of values for \(\Omega\) and \(\gamma_2\), the two models are in good agreement, thus justifying the approximation of reducing the four-level system to a three-level system.

\section{\label{section:strong incoherent pumping} Optimizing cooling in the presence of strong incoherent pump}
\begin{figure}[h]

    \centering
    \includegraphics[width = 0.30\textwidth]{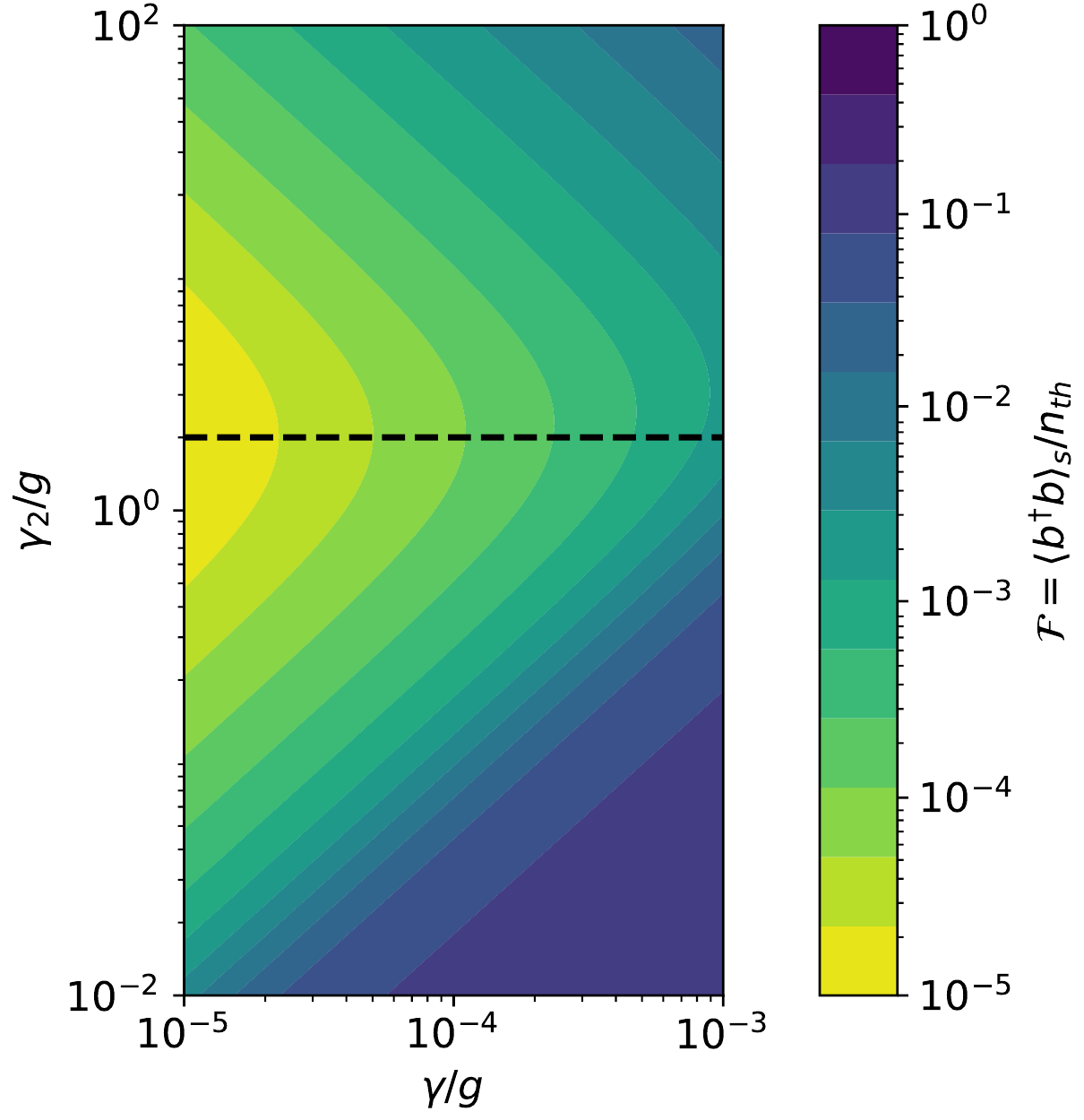}
    \caption{Variation of \(\mathcal{F}\) with the decay rate of second excited state \(\gamma_2\) and phonon decay rate \(\gamma\) under strong incoherent pump. The black dashed line represents the optimal decay rate as obtained from our model. Here, \(\omega_m = 200\) MHz, \(n_{th} \approx 200\), \(g = 1\) MHz, \(\gamma_p = 10g\), \(\gamma_1 = 10^{-5}g\), and the cutoff for the Fock state basis \(N = 300\).}
    \label{incoherent_optimal}
\end{figure}

Following the reference \cite{Cortes_2019}, we set \(\omega_m = 200\) MHz, \(n_{th} \approx 200\), \(g = 1\) MHz, and assume \(\gamma_1 = 10^{-5}g\) and \(\gamma_p = 10g\). Using the quantum toolbox, we plot the variation of \(\mathcal{F}\) as a function of the phonon decay rate \(\gamma\) and the decay rate of the second excited state \(\gamma_2\) in Figure \ref{incoherent_optimal}. The black dashed line represents the line \(\gamma_2 = 2g\), the optimal decay rate obtained from analytical calculations. The results are in good agreement with the results obtained in \cite{Cortes_2019} for the case of a single atom. 

%\bibliography{biblography}

\providecommand{\noopsort}[1]{}\providecommand{\singleletter}[1]{#1}%

\end{document}